\pdfoutput=1
\documentclass[pre,aps,twocolumn,showpacs, showkeys,floatfix,nofootinbib,
superscriptaddress
]{revtex4-1}
\usepackage{subfigure}
\usepackage{amssymb}
\usepackage{amsfonts}
\usepackage{amsmath}
\usepackage{amsthm}
\usepackage{epsfig}
\usepackage{graphicx}
\usepackage[usenames,dvipsnames]{color}
\usepackage[latin1]{inputenc}
\usepackage{tabu}

\usepackage[hidelinks,unicode=true]{hyperref}
\hypersetup{colorlinks=true,
 	linkcolor=blue,
 	urlcolor=blue,
 	citecolor=blue,
 	pdfhighlight=/N
 }
 \usepackage{comment}
 \usepackage[normalem]{ulem}

\renewcommand{\vec}{\mathbf}

\begin{document}

\title{Effects of a kinetic barrier on limited-mobility interface growth models}

\author{Anderson J. Pereira}
\affiliation{Departamento de F\'{\i}sica, Universidade Federal de
	Vi\c{c}osa, Minas Gerais, 36570-900, Vi\c{c}osa, Brazil}
\author{Sidiney G. Alves}
\email{sidiney@ufsj.edu.br}
\address{Departamento de Estatística, F\'isica e Matem\'atica, Campus Alto Paraopeba, Universidade Federal de S\~ao Jo\~ao Del-Rei, 36420-000, Ouro Branco, MG, Brazil}
\author{Silvio C. Ferreira}
\affiliation{Departamento de F\'{\i}sica, Universidade Federal de
 Vi\c{c}osa, Minas Gerais, 36570-900, Vi\c{c}osa, Brazil}
\affiliation{National Institute of Science and Technology for Complex Systems, 22290-180, Rio de Janeiro, Brazil}

\begin{abstract}
The role played by a kinetic barrier originated by out-of-plane step edge
diffusion,   introduced in [Leal \textit{et al.},
\href{https://doi.org/10.1088/0953-8984/23/29/292201}{J. Phys. Condens. Matter
	\textbf{23}, 292201 (2011)}], is investigated in the Wolf-Villain and Das
Sarma-Tamborenea models with short range diffusion. Using large-scale
simulations, we observe that this barrier is sufficient to produce growth
instability, forming quasiregular mounds in one and two dimensions. The
characteristic surface length saturates quickly  indicating a uncorrelated
growth of the three-dimensional structures, which is also confirmed by a growth exponent
$\beta=1/2$. The out-of-plane particle current  shows a large reduction of
the downward flux in the presence of the kinetic barrier enhancing, consequently, the
net upward diffusion and the formation of three-dimensional self-assembled structures.
\end{abstract}

\maketitle

\section{Introduction}

A rich variety of morphologies can be observed during far-from-equilibrium
growth processes and many of them with potential for technological
applications~\cite{michely2004islands,Evans2006,barabasi,meakin}.  Growth
instability can induce three-dimensional mound-like patterns  in different types
of films such as metals \cite{Jorritsma,Caspersen,Han},  inorganic
\cite{Johnson,Tadayyon} and organic \cite{Zorba,Hlawacek} semiconductors
materials to cite only a few examples. Such a growth instability has been mainly
attributed to the presence of  Ehrlich-Schwoebel (ES) step barriers
\cite{Ehrlich,Schwoebel} that reduce the rate with which atoms move downwardly
on the edges of terraces leading to net uphill flows. Growth instabilities can
also emerge from topologically induced uphill currents which depend on the
crystalline structure \cite{Kanjanaput2010} or from fast diffusion on terrace
edges~\cite{Murty2003,Pierre-Louis1999} among other
mechanisms~\cite{Evans2006,michely2004islands}. The existence of ES barriers is
supported by molecular dynamic simulations~\cite{Yang}.

\begin{figure*}[tp]
	\centering
	\includegraphics[width=0.35\linewidth]{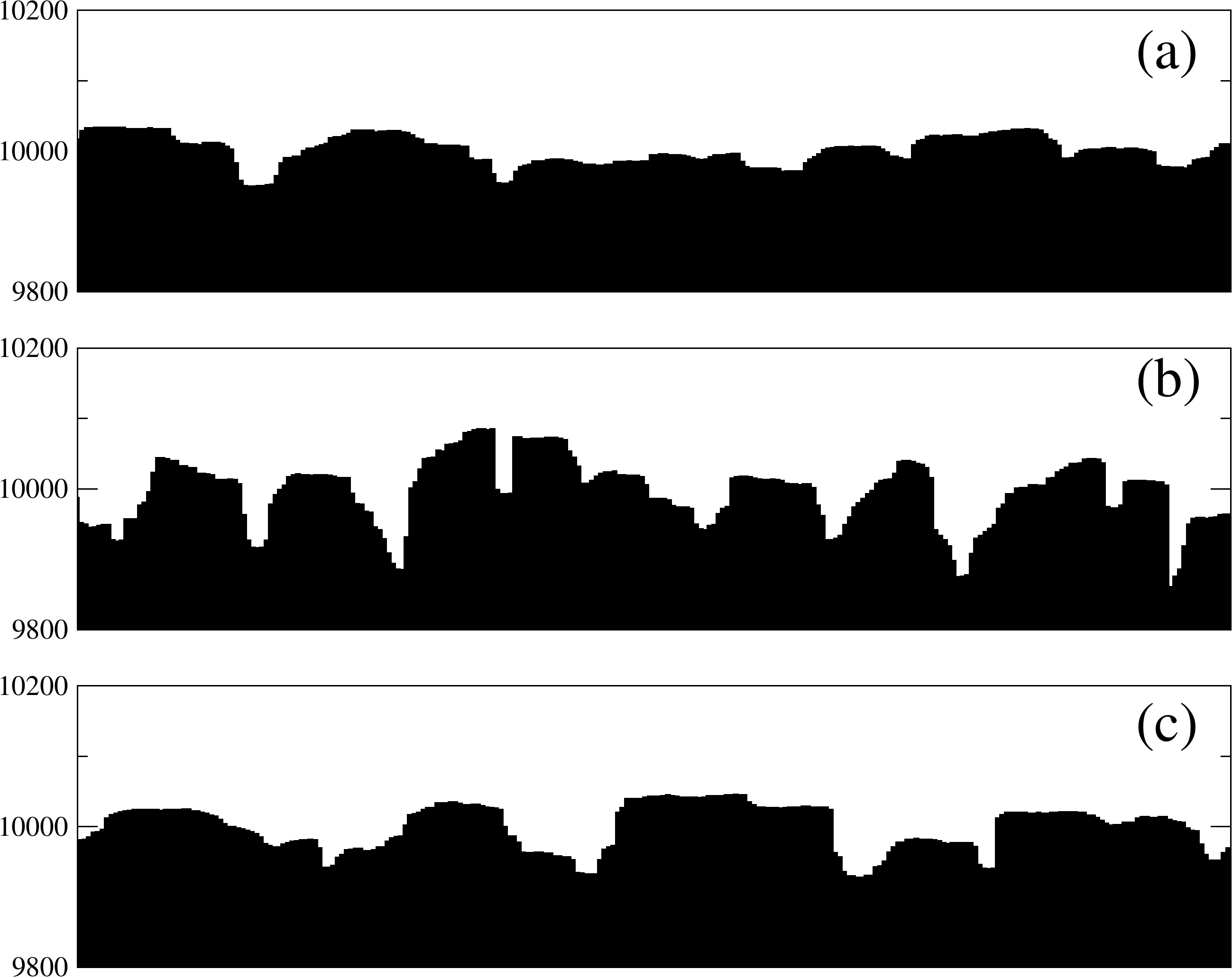} ~~~
	\includegraphics[width=0.35\linewidth]{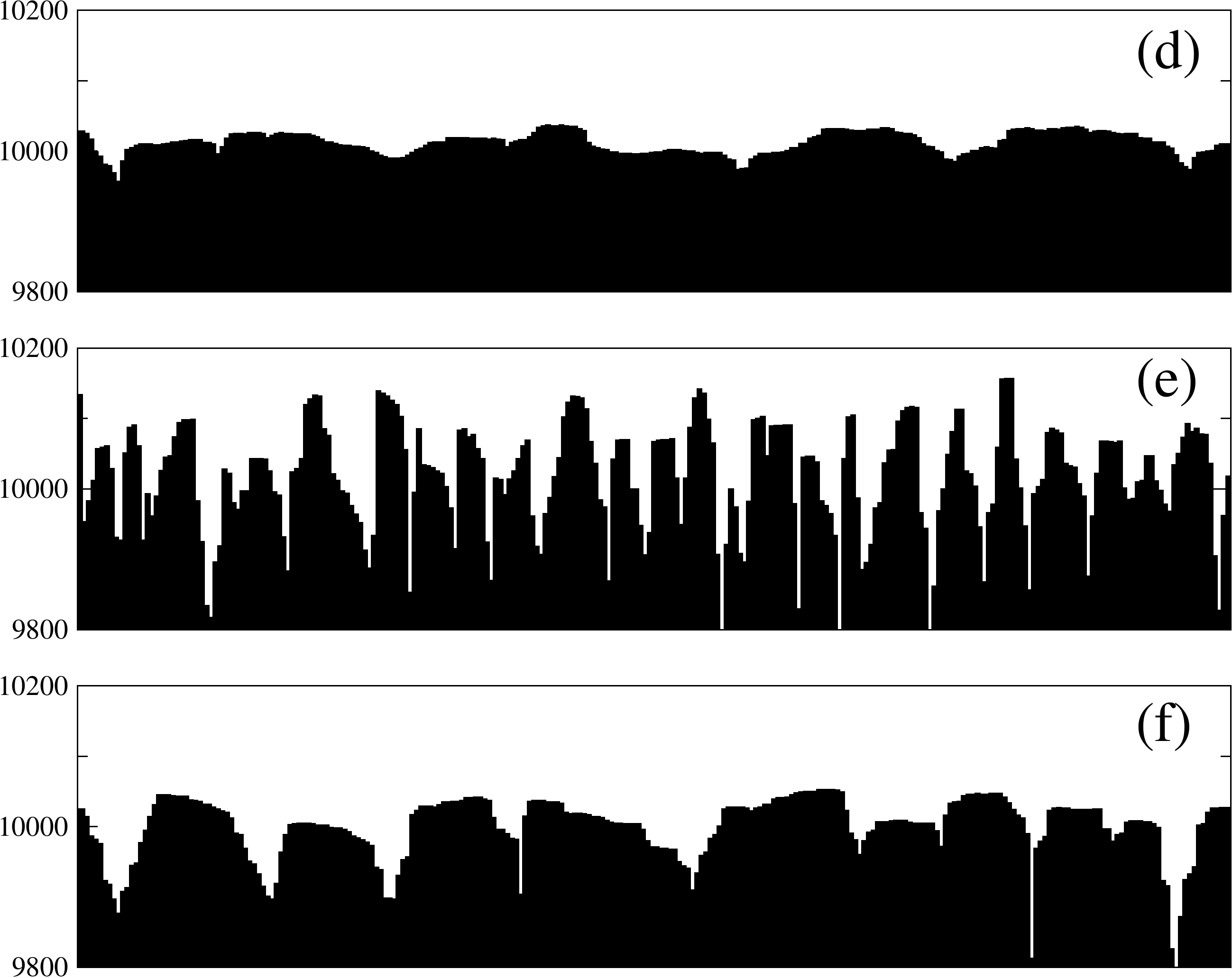}
	\caption{\label{snapshot_1d} Interfaces of the WV and DT models in $d=1$ shown
		in left and right panels, respectively. Cases (a,d) without and with the
		kinetic barrier considering (b,e) $N_s=1$  and (c,f) $N_s=10$ are shown. All the
		simulations were done on a lattice of size $L=2^{10}$ for a deposition time $t=10^4$.}
\end{figure*}

Discrete solid-on-solid (SOS) growth models constitute an important approach to
investigate the dynamic of kinetic roughening and morphological properties of
interfaces. The rules are easily implemented in a discrete space (lattices)
rid of overhangs and bulk voids. The role played by ES barriers has been
investigated in models with thermally activated
diffusion~\cite{Evans2006,michely2004islands} being the Clark-Vvedenski (CV)
model~\cite{CV_PRL,Clarke1988} one of the simplest examples, in which any surface
adatom can move according to an Arrhenius diffusion coefficient $D\sim
\exp(-E/k_B T)$~\cite{barabasi} where $E$ is an energy activation barrier to be
overcome in a diffusion hopping. An ES barrier can be included as an additional activation
energy for diffusion at the edges of terraces~\cite{Evans2006}. The effects of a step barrier of
purely kinetic origin, namely simple diffusion, were investigated in an epitaxial growth model with
thermally activated diffusion~\cite{Leal_JPCM}. In this model,  a particle
performing an interlayer movement through steps with more than one monolayer has
to diffuse along the columns, perpendicularly to the substrate, instead of
attaching directly at the bottom or top of a terrace. This kinetic barrier reduces
downhill currents and three-dimensional structures in the form of mounds are
obtained at short-time scales even in the case of weak ES barriers where the
conventional rule would not lead to mound formation.

Simple models with limited mobility can be used to investigate kinetic
roughening~\cite{barabasi,meakin}. Wolf-Villain (WV)~\cite{WV} and Das
Sarma-Tamborenea (DT)~\cite{DT} models, introduced to investigate
molecular-beam-epitaxy (MBE) growth, are benchmarks of this class and  have been
intensively investigated~\cite{Smilauer, Milan, Huang,
	HaselwandterPRL,HaselwandterPRE,Sarma,Punyindu,wvbogo,Xun,Luis2019}. A variation
of the CV model with limited mobility has been
considered~\cite{Aarao2010,Aarao2013} and many features of the original model
have been reproduced with this simplified version~\cite{To2018}. Effects of a
step barrier were investigated in both WV~\cite{Rangdee} and DT
\cite{DasSarma_SC} models introducing two additional probabilities for downward
and upward interlayer diffusion with the former larger than the latter, and
mound formation was observed in both models. WV and DT models without step
barrier were investigated in several
lattices~\cite{Chatraphorn2001,Kanjanaput2010} and it was found that the WV
model can present topologically induced mound morphologies on some lattices but
not in others while no clear evidence for three-dimensional structures was
observed for DT. In one-dimension, it is widely accepted that both DT and WV
models asymptotically produce self-affine surfaces belonging to nonlinear
MBE~\cite{Luis2019} and Edwards-Wilkinson~\cite{Vvedensky} universality classes,
respectively.

It was reported that a kinetic barrier alone does not induce mound morphologies
in thermally activated  CV-like models~\cite{Leal_JPCM} but, instead, they
exhibit kinetic roughening with exponents consistent with the nonlinear MBE
universality class~\cite{DT,Villain,LSarma}. Therefore, given the simplicity of
limited-mobility growth models and the non-trivial effects of topologically
induced uphill currents in DT and WV models, one would wonder how they respond
to a barrier of purely kinetic origin. In order to fill this gap, we investigate
WV and DT models with the introduction of the kinetic barrier proposed in
Ref.~\cite{Leal_JPCM}. We observed mounds in  both models in 1+1 and 2+1
dimensions, being much more evident for WV model. The surface coarsening ceases
quickly with the saturation of the characteristic surface length and regimes of
uncorrelated mound growth are asymptotically observed. Analysis of the
out-of-plane currents shows a large reduction of the downhill flux of particles,
enhancing surface instabilities and mound formation.

The remaining of the paper is organized as follows. The model implementation details
are presented in section~\ref{sec:model}. In section \ref{results}, we discuss
the results obtained in the simulations. Our conclusions and some perspectives are drawn in the
section~\ref{conclusion}.

\section{Models}
\label{sec:model}

In all investigated models, the particles are randomly deposited on a
$d$-dimensional lattice of linear size $L$ with periodic boundary conditions
under the SOS condition. Results presented in this work correspond to regular chains
in $d=1$ and square lattices in $d=2$. Other lattices were tested and the
central conclusions remain unaltered. The height of the interface at site $i$ and
time $t$ is represented by $h_i(t)$ and the initial condition is given by
$h_i(0)=0$  such that the initial interface is flat.

In the WV model with a kinetic barrier {investigated in the present work},
the growth rule is implemented as follows. At each time step, a position $i$ is
randomly chosen.
A location $i'$ with the largest number of
bonds that a new deposited adatom would have is determined within a set
containing $i$ and its nearest-neighbors. If the initial position corresponds to
the largest number of bonds ($i'\equiv i$), it is chosen as the deposition place
and the simulation runs to the next step.  In case of multiple  options, one is chosen at
random. Otherwise, the particle tries to diffuse
to the neighbor $i'$ with a probability given by~\cite{Leal_JPCM}
\begin{equation}
\label{prob}
P_{\delta h}(i,i')=
\left\{
\begin{array}{cl}
1, & \textrm{if } ~ |\delta h|<2\\ 
\frac {1}{|\delta h|}, & \textrm{if } ~ |\delta h|\geq 2 
\end{array}
\right.
\end{equation}
where $\delta h = h_i-h_{i'}$.  With probability $1 - P_{\delta h}(i,i')$ the particle remains at the
site $i$. It is important to mention that Eq.~\eqref{prob} is obtained {
	assuming that the adatom first moves to top kink of the terrace and then start a
	unbiased one-dimensional random-walk normally to the initial substrate, stopping
	the movement if it either arrives at the bottom or return to top of the terrace.
	The result is the} solution of a non-directed one-dimensional random walk  with
absorbing  boundaries separated by a distance $|\delta h|$
\cite{Shehawey}; see Fig.~1 of Ref.~\cite{Leal_JPCM} for further details of this diffusion
rule. This diffusion attempt is successively applied $N_s$ times (representing a
$N_s$ diffusive steps) departing from the last position of the adatom. A unit
time is defined  as the deposition of $L^d$ particles.

The implementation of the DT model with kinetic barrier is similar. The
difference is that diffusion to the nearest-neighbors are performed only if the
adatom does not have  lateral bounds  and  any neighbor with a number of bonds
higher than 1  can be chosen with equal chance as the target site.

\begin{figure*}[ht]
	\centering
	\includegraphics[width=0.32\linewidth]{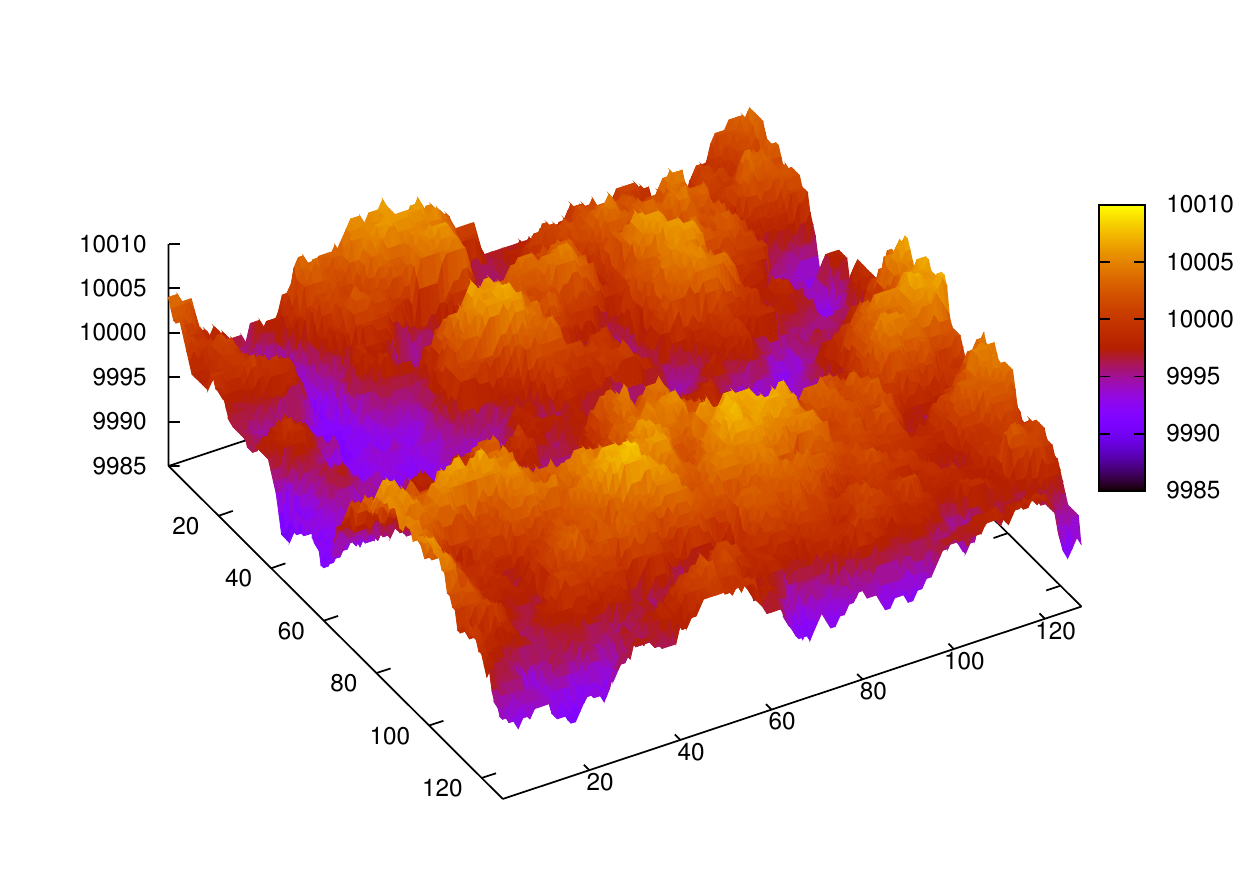}
	\includegraphics[width=0.32\linewidth]{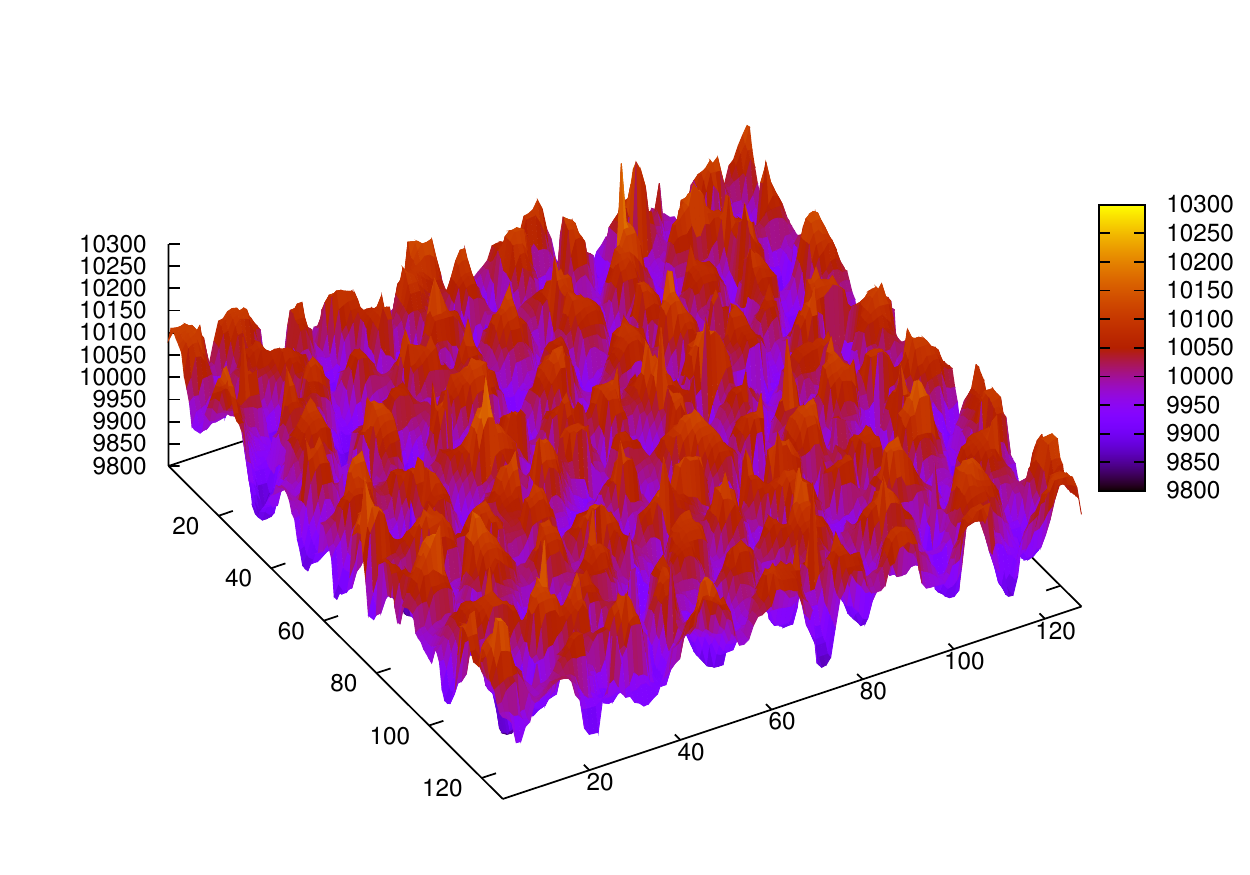}
	\includegraphics[width=0.32\linewidth]{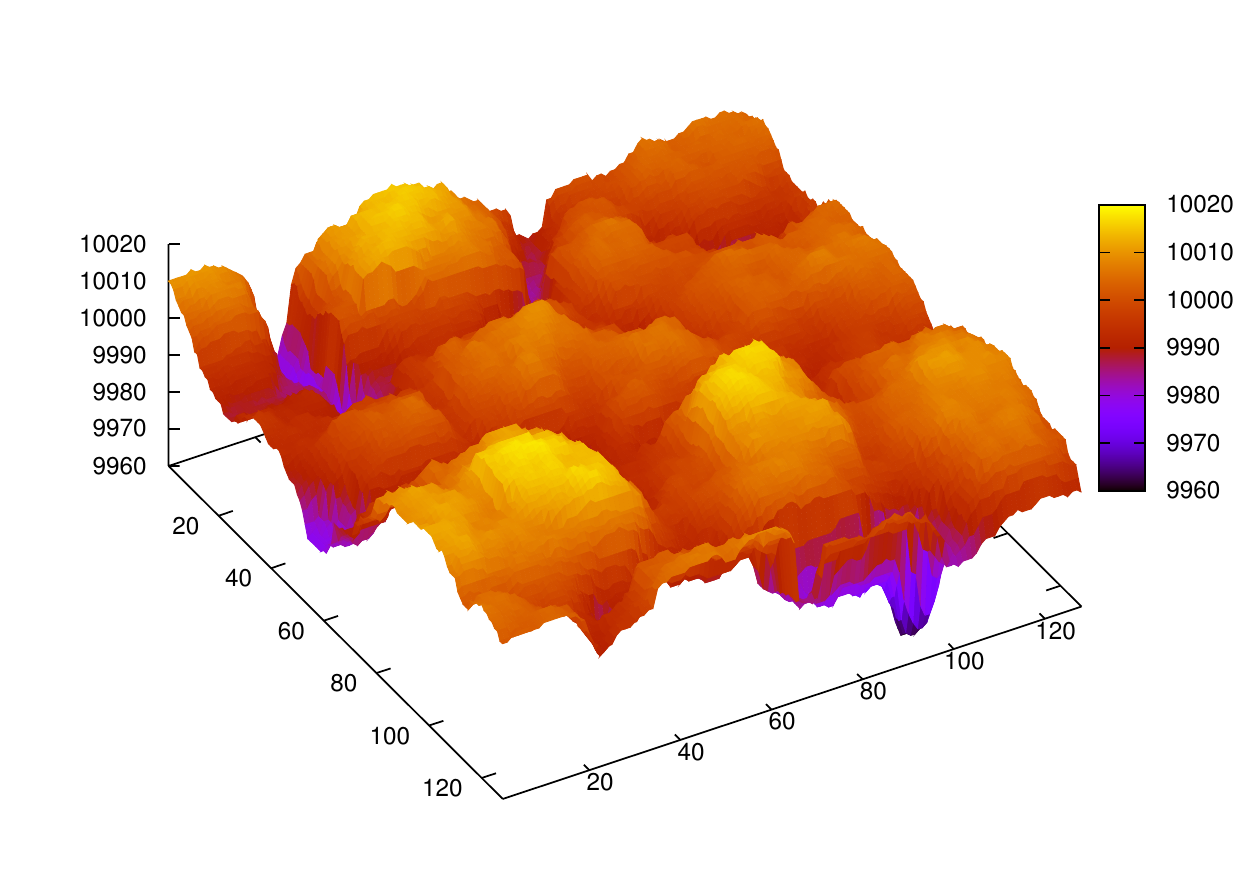}\\
	\includegraphics[width=0.32\linewidth]{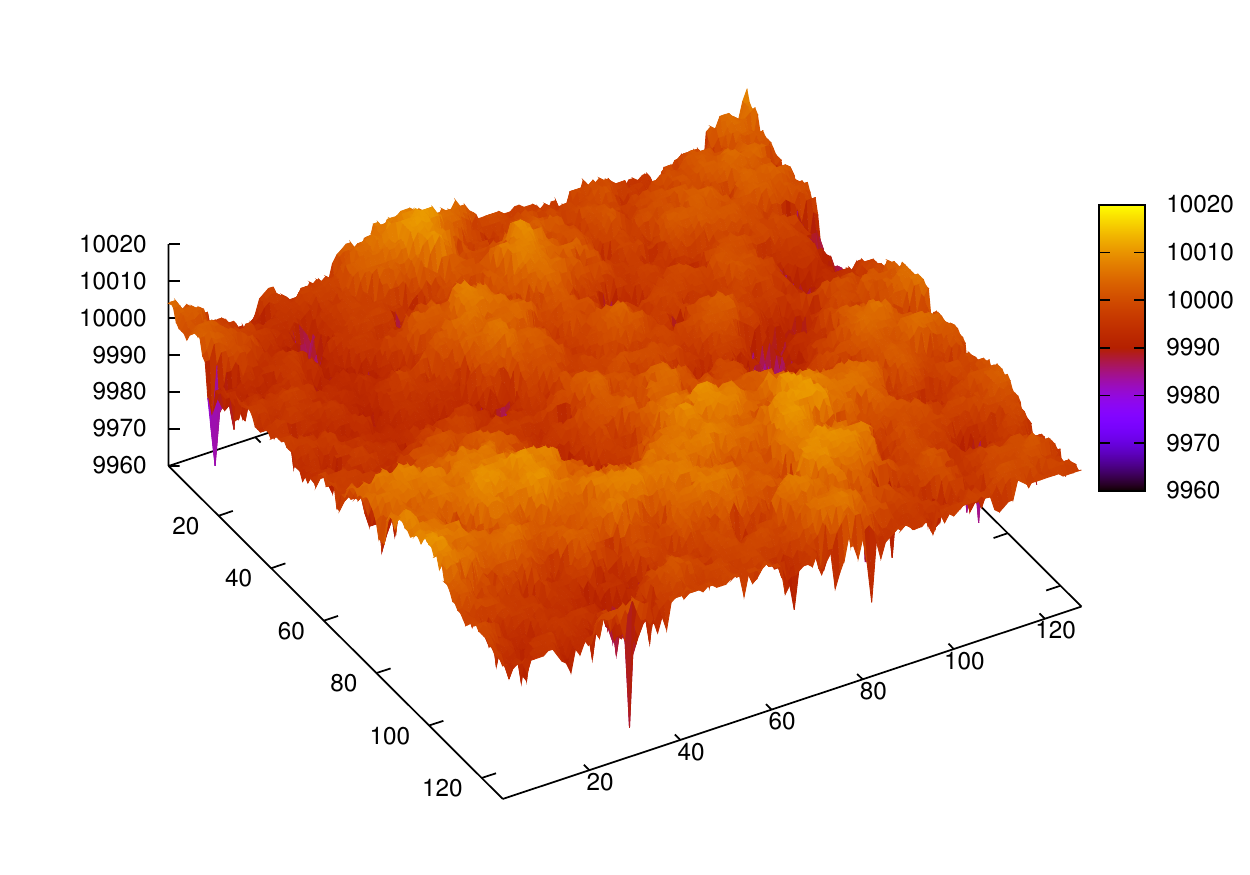}
	\includegraphics[width=0.32\linewidth]{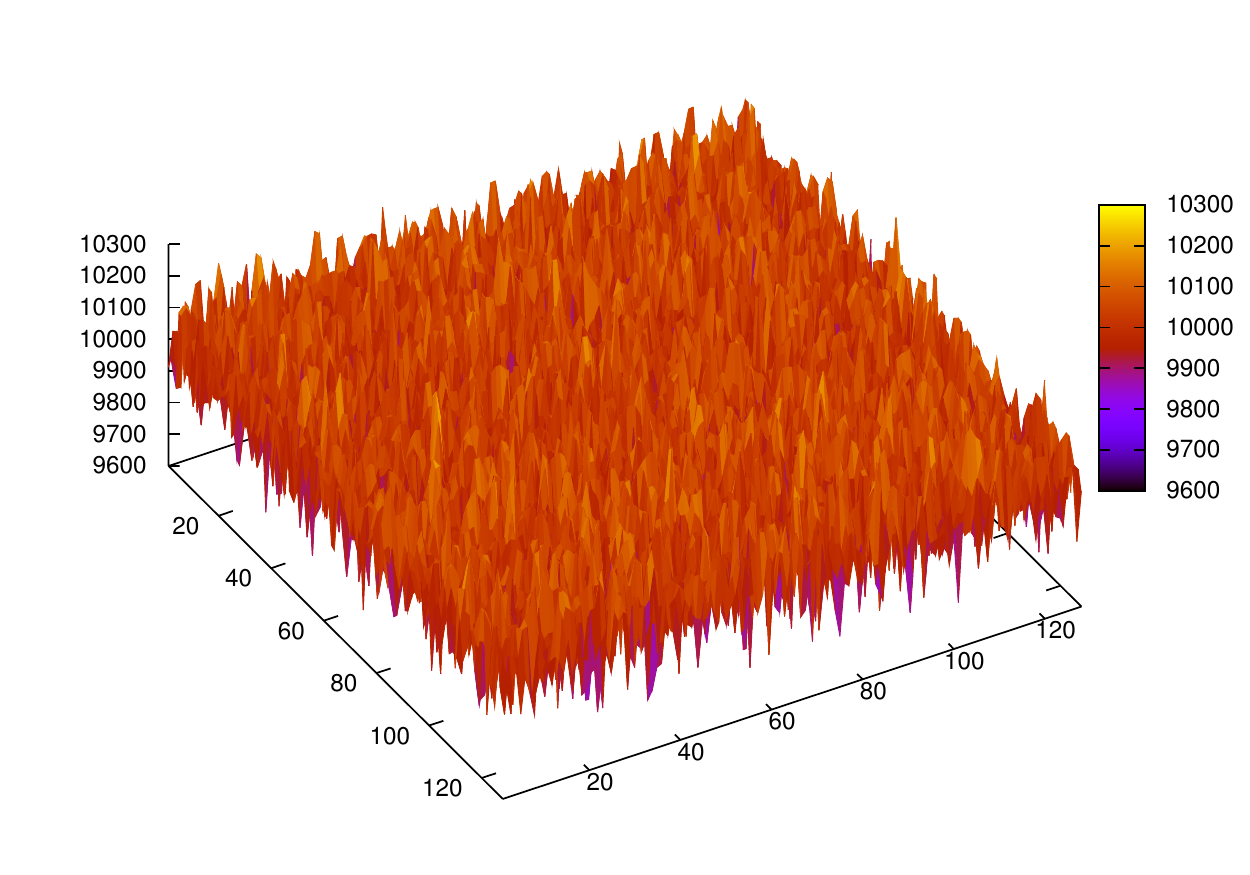}
	\includegraphics[width=0.32\linewidth]{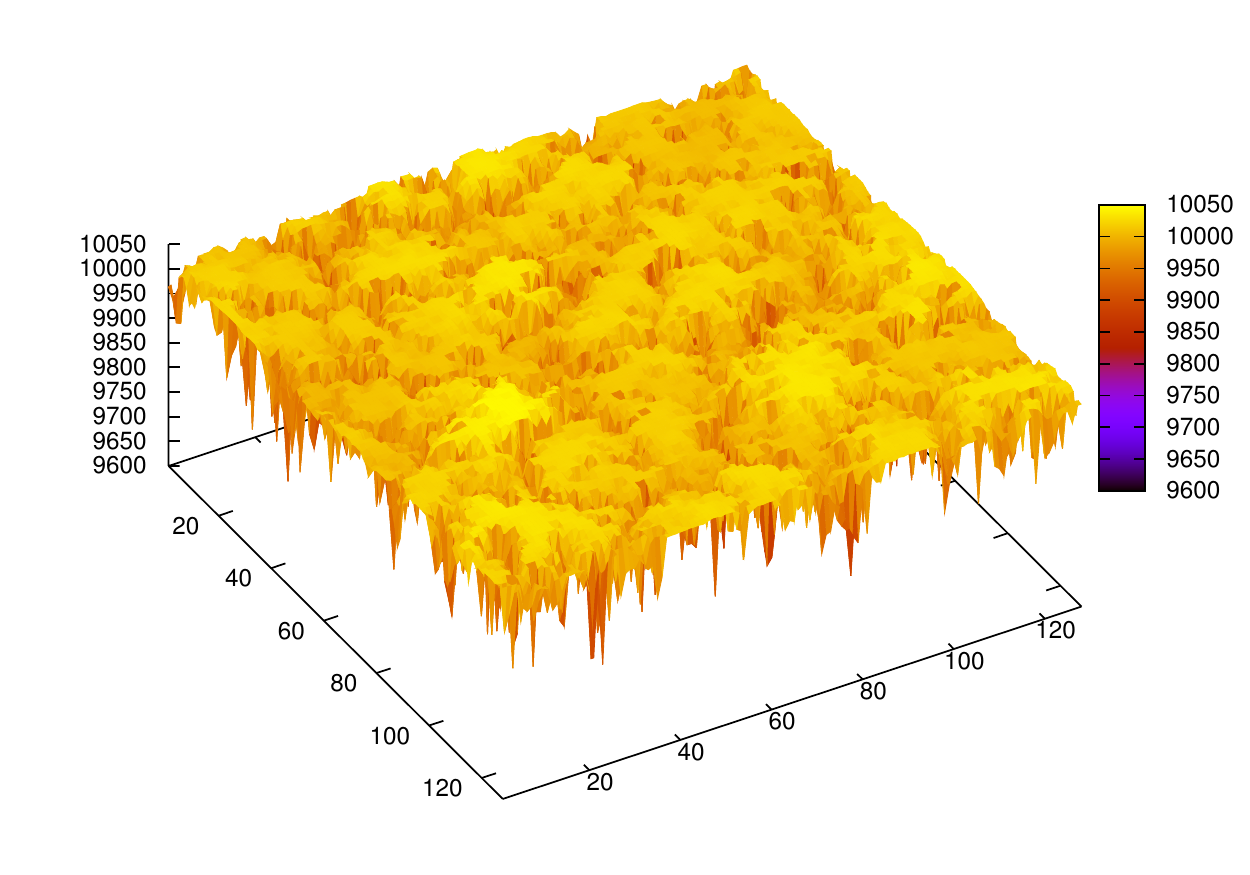}
	\caption{\label{snapshot_2d} Interfaces obtained using the WV and DT models in
		$d=2+1$ are shown in top and  bottom panels, respectively. The case without
		(left) and with the kinetic barrier  considering $N_s=1$ (center)  and $N_s=10$
		(right) are shown.  All simulations were done on square lattices of size $L=2^{9}$
		and a deposition time $t=10^4$.}
\end{figure*}

\begin{figure}[htp]
	\centering
	\includegraphics[width=0.8\linewidth]{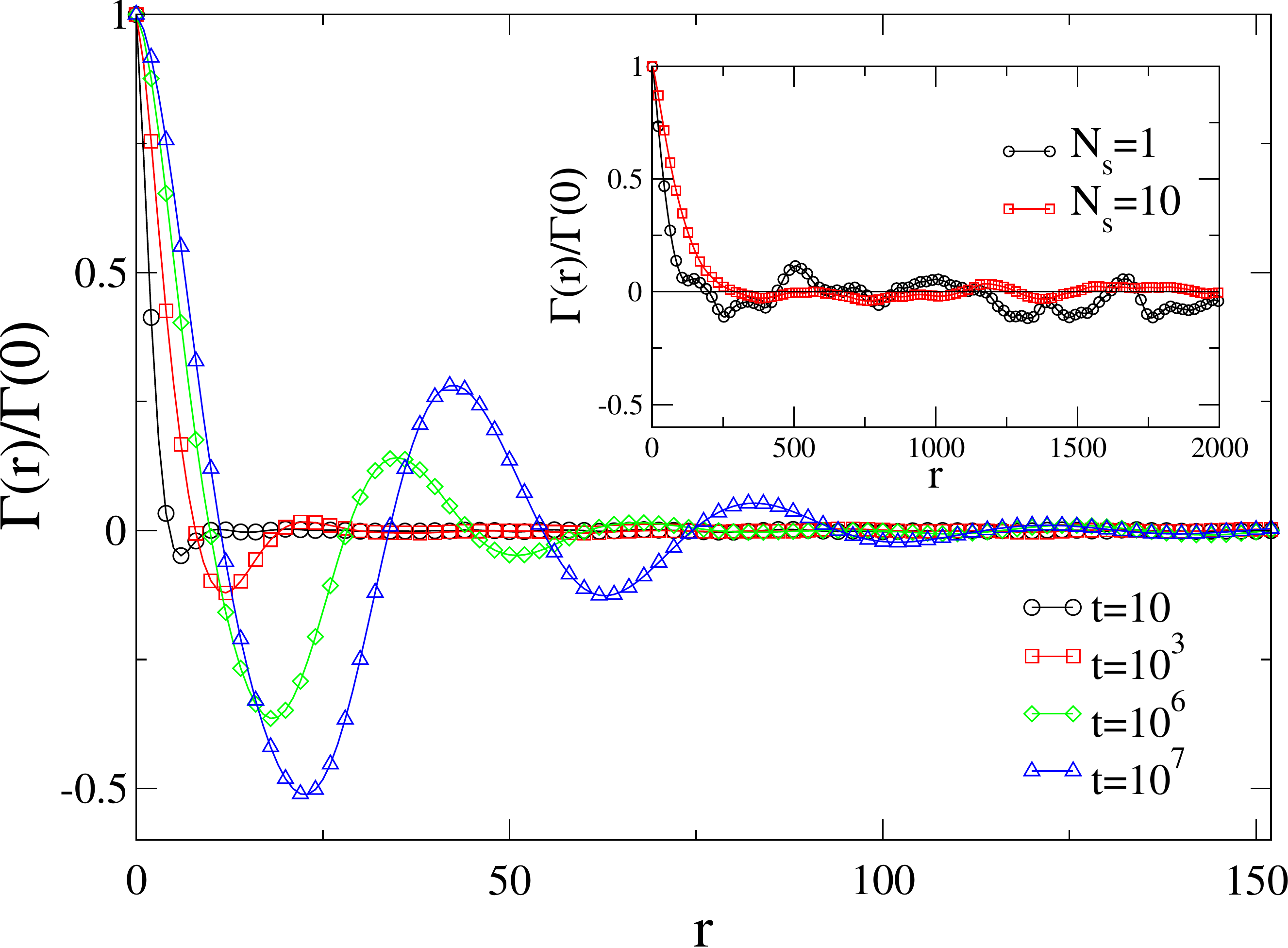}\\
	\includegraphics[width=0.8\linewidth]{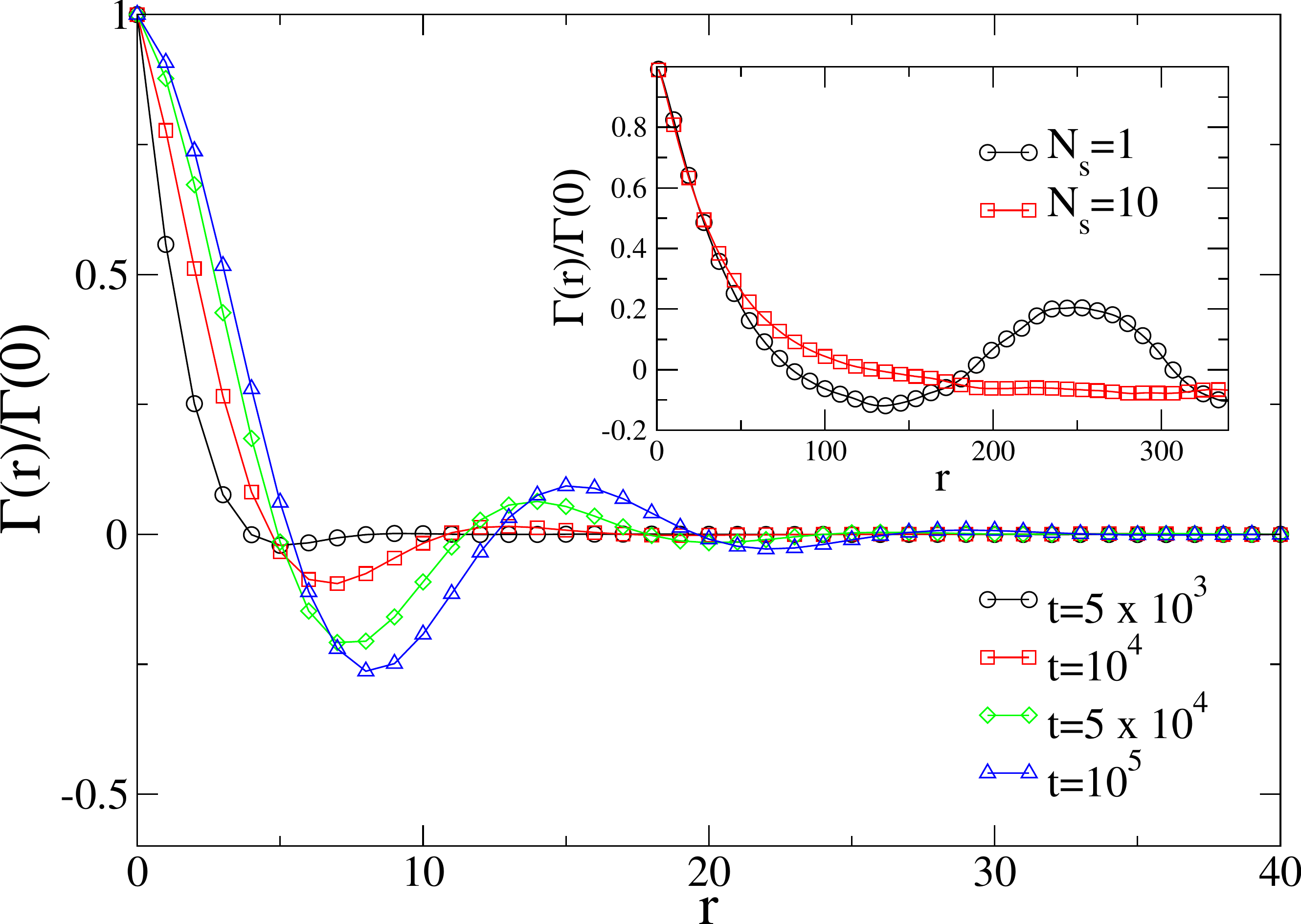}
	\caption{\label{hh_corr_time}Main panels: Height-height correlation function
		for the WV model at distinct times indicated in the legends for (a) one- 
		and (b) two-dimensional substrates. The number of steps is $N_s=1$. The
		averages were computed  over 100 independent runs. Insets: Correlation
		functions averaged over 1 and 10 samples for the original WV model at time
		$t=10^5$ showing that the oscillations observed in single samples are not due
		to regular structures.}
\end{figure}
\section{Results}
\label{results}

The one-dimensional simulations were carried out on chains with up to $L=2^{14}$
sites and evolution times of up to $t = 10^7$. In the
two-dimensional case, the simulations were done in systems of size up to
$L=2^{10}$ and time up to $t = 10^6$. The averages were performed over $100$
independent runs.

Figures~\ref{snapshot_1d} and \ref{snapshot_2d} show interfaces obtained in
simulations in one- and two-dimensional substrates, respectively.
Surfaces for the original WV and DT models without  and with ($N_s=1$
or $N_s=10$)  kinetic barriers are compared. In both dimensions, the irregular
morphologies without a characteristic length observed in the original versions
change to structures separated by valleys  that present a well-defined
characteristic length. We also observe that an increase in the value of $N_s$
reduces valley deepness and increases  the characteristic width of the mounds. The
effects of the kinetic barrier seem to be stronger in two- than one-dimension. A
remarkable change in the profiles happens when just one hop to nearest-neighbors
is allowed in the DT model with kinetic barrier, as can been seen in
Fig.~\ref{snapshot_1d}(e). Surfaces become columnar with a high aspect ratio
(height/width). Such a behavior is reminiscent of the very strict rule for
diffusion in DT when a single lateral bound is enough to irreversibly stick the
adatom on a site. In the WV case, where diffusion happens more readily, mound
morphologies with quasiregular structures emerge more clearly.

\begin{figure}[ht]
	\centering
	\includegraphics[width=0.8\linewidth]{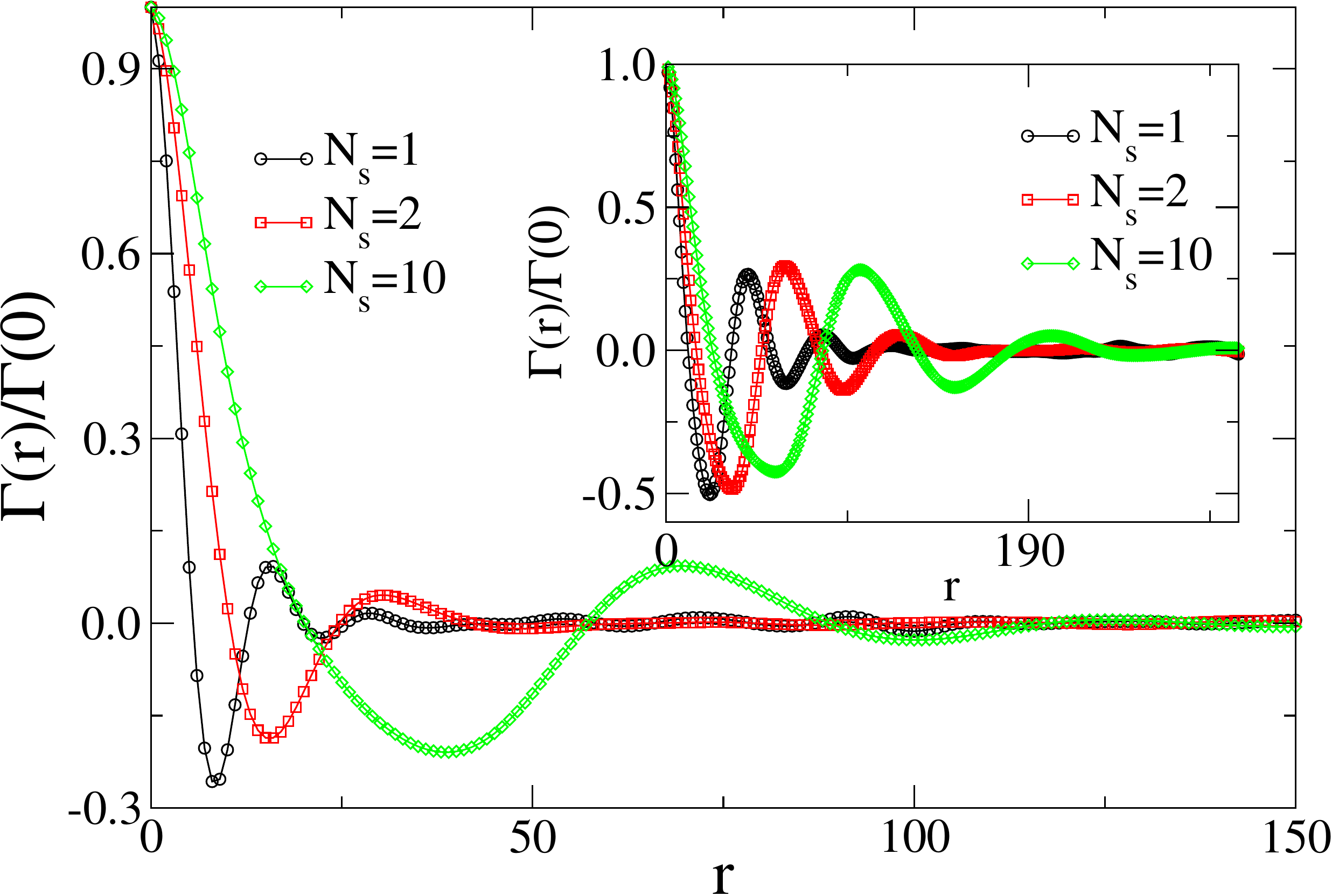}
	\caption{\label{hh_corr_Ns}
		Main plot: Height-height correlation function dependence with the parameter
		$N_s$ (indicated in the legends) for the WV model in two-dimensional
		substrates at a time $t=10^5$. Inset: Same as the main plot for  one dimension
		at a time $t=10^7$.  Curves correspond to averages
		over 100 independent samples.}
\end{figure}

A standard tool to characterize the
morphology of interfaces in growth process is the height-height correlation
function defined as \cite{Evans2006,Murty2003,Chatraphorn2001}
\begin{equation}
\Gamma(\vec r) = \left\langle \tilde h(\vec x) \tilde h(\vec x + \vec r) \right\rangle_x, 
\end{equation}
here $\tilde h(\vec x)$ is the height interface at position $\vec x$ relative to
the mean height and $\left\langle \ldots \right\rangle_x$ denotes an average
over the surface. The height-height correlation for $\vec r = 0$ is related to
the interface width by
\begin{equation}
\sqrt{\left\langle\Gamma(0)\right\rangle} = w
\end{equation}
here $\left\langle\ldots\right\rangle$ denotes an average over independent runs.
A self-affine interface is characterized by a height-height correlation function
that goes monotonically to zero while those characterized by mounds exhibit
oscillatory behavior around $0$. In the latter case, the first zero of
$\Gamma(r)$, denoted by $\xi$, is a characteristic lateral length of the
mounds in the surface.

Figure~\ref{hh_corr_time} shows the height-height correlation function for the WV
model with kinetic barrier in one- and two-dimensional substrates. The curves
clearly exhibit oscillatory behavior  even for averages over 100 independent
samples. Conversely, the irregular oscillatory behavior observed for the original  WV model
 shown in insets of Fig.~\ref{hh_corr_time} is
lumped after averaging. Therefore, interfaces obtained with kinetic
barrier are characterized by the formation of quasiregular mound structures
differently from those obtained using the original model that exhibits
irregular structures within the intervals of size and time  we
investigated. These plots also show a coarsening of the mounds represented by
the first minimum displacement at the early growth times. 

\begin{figure}[ht]
	\centering
	\includegraphics[width=0.8\linewidth]{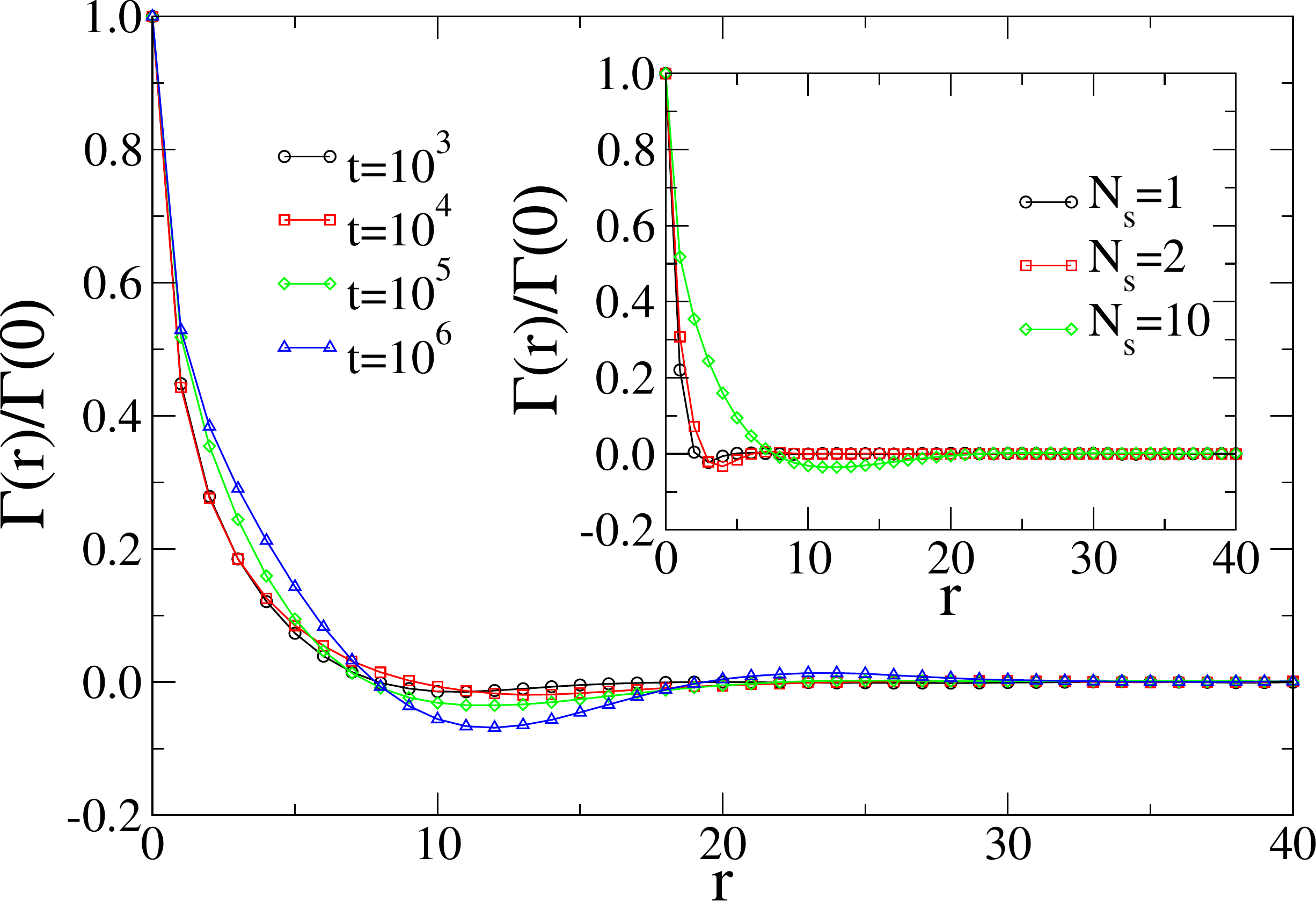}
	\caption{Main plot: Correlation function for the DT model in two-dimensional
		substrates for distinct times shown in the legends and fixed $N_s=10$. Inset:
		Correlation function for DT model in two dimensions at a fixed time $t=10^5$
		and different values of $N_s$ shown in legends. Curves correspond to averages
		over 100 independent samples.}
	\label{h_corr_DT}
\end{figure}

The effect of the parameter $N_s$ in WV model is shown in Fig.~\ref{hh_corr_Ns}.
As indicated by the interface profiles shown in Figs.~\ref{snapshot_1d} and
\ref{snapshot_2d},  the characteristic lateral length increases with
$N_s$ in both dimensions.
The correlation function for DT model follows a qualitative similar dependence
with $N_s$, as can be seen in Fig.~\ref{h_corr_DT} where the effects of time and
number of diffusion steps in the correlation function of the DT model are shown.
However, the mounds are much less evident than those obtained in the WV model.
However,  the correlation functions still present the typical oscillatory
behavior of mounded structures that is preserved after the averaging over
100 independent samples. Besides,  the typical width of the mounds in the DT
model are much smaller than those of WV. It is important to note that the
correlation function of the  original DT model also presents an irregular
behavior as does the WV model.

\begin{figure}[hbt]
	\centering
	\includegraphics[width=0.8\linewidth]{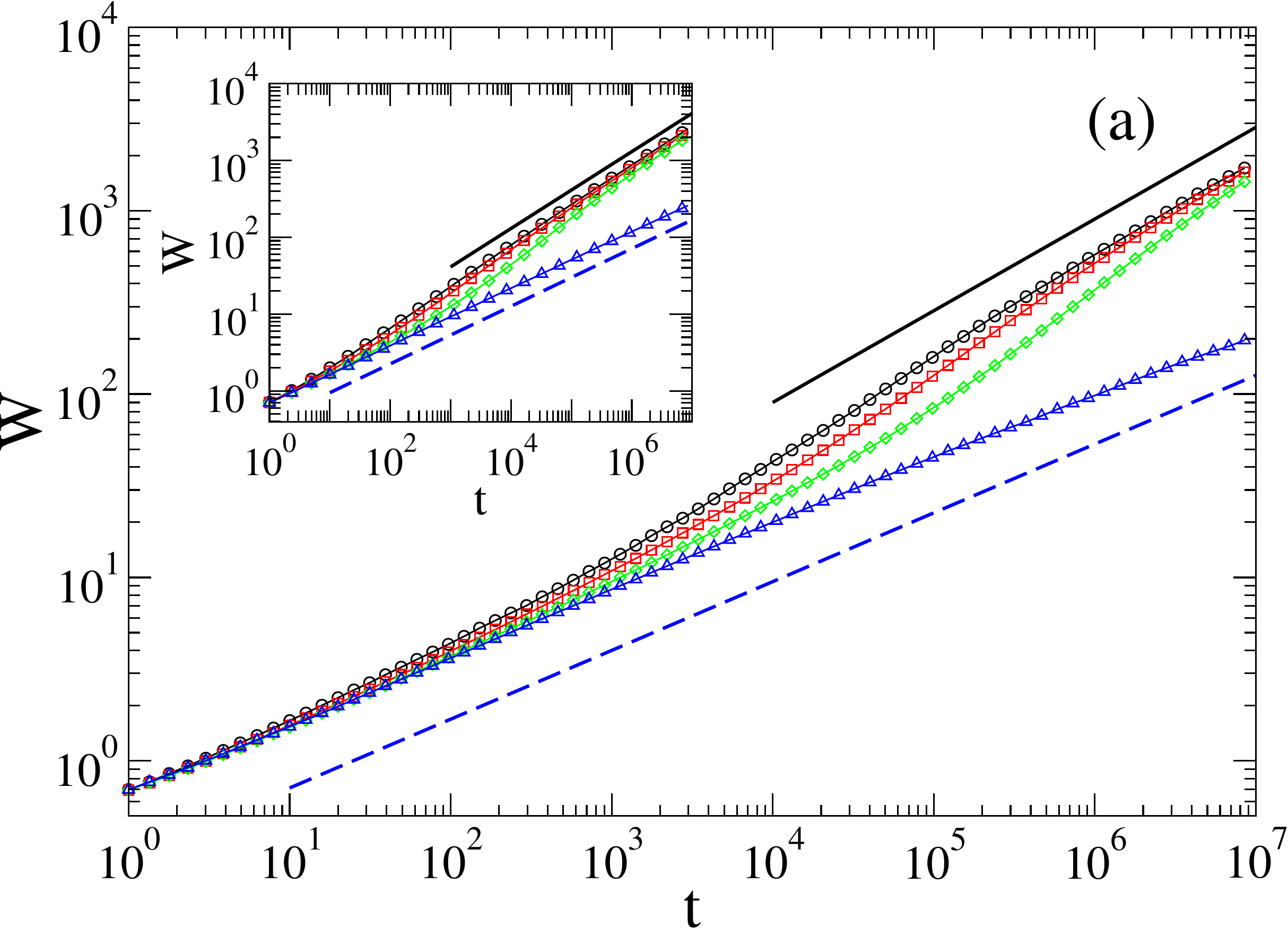}\\
	\includegraphics[width=0.8\linewidth]{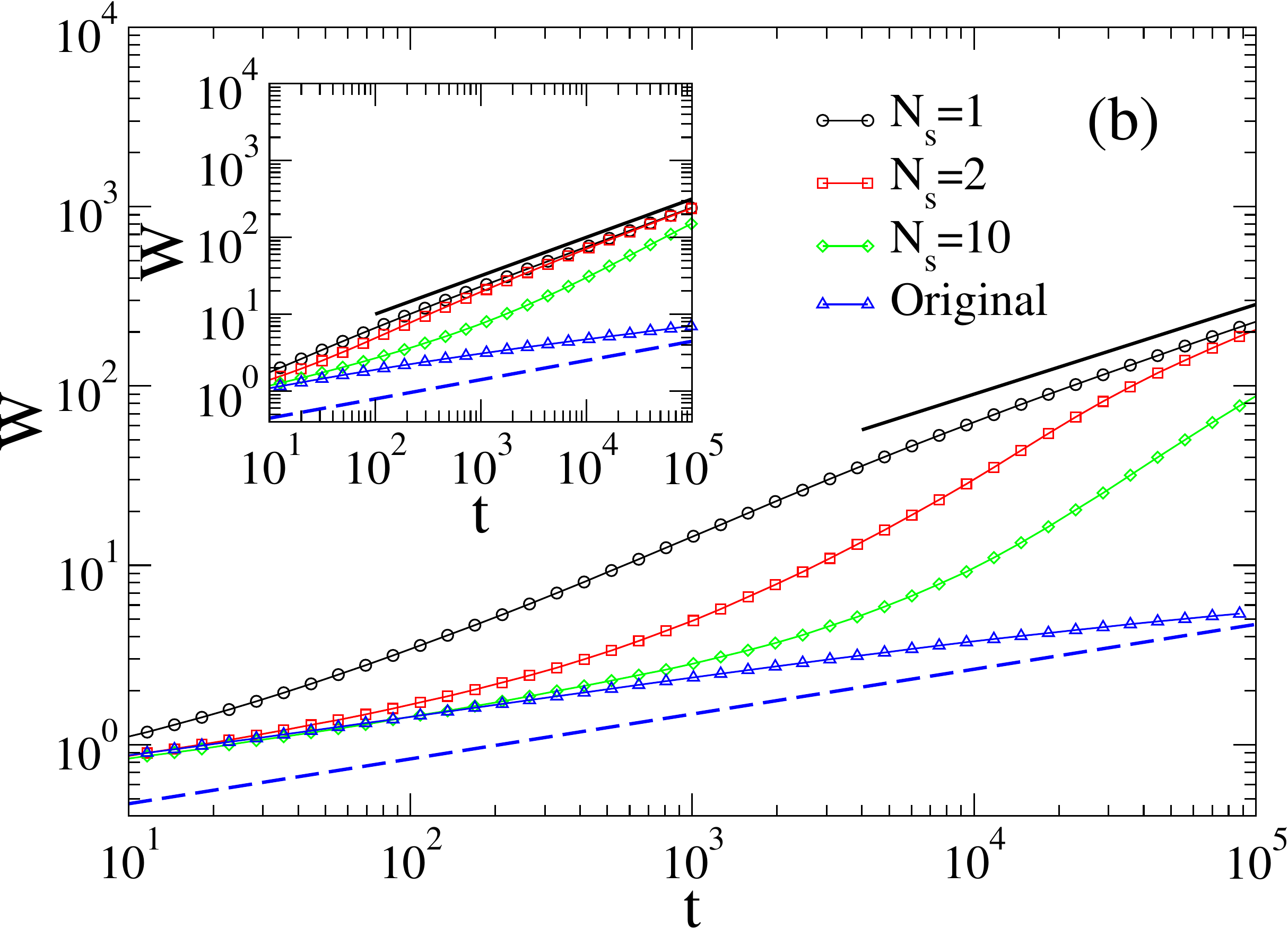}
	\caption{\label{width}Time evolution of the  interface width  $w$ for
		WV (main panels) and DT (insets) models grown on (a) one- and (b)
		two-dimensional substrates. Both  simulations with  the kinetic
		barrier (using $N_s$ values indicated in the legend) and the original
		version are shown. In (a), dashed  and solid lines are power-laws with exponents 
		$3/8$ and $1/2$, respectively, in both main panels and insets. In (b),
		the slopes of the dashed and solid lines are $1/4$  and $1/2$, respectively.}
\end{figure}
Figure~\ref{width} shows the time evolution of the interface width for both models
in one and two dimensions. The main panels and insets present the
results for the WV and DT models, respectively, including  or not the kinetic
barrier.  The interface width is expected to scale as $w\sim t^\beta$ where
$\beta$ is the growth exponent~\cite{barabasi}. The short time dynamics of both
WV and DT models  is well described by the linear version of
the MBE equation~\cite{Villain,LSarma}
\begin{equation}
\frac{\partial h}{\partial t} = -\nu \nabla^{4}h + \lambda \nabla^{2} (\nabla h)^{2} + \eta,
\label{Eq1}
\end{equation}
with $\lambda=0$ where $\eta$ is a non-conservative Gaussian
noise~\cite{Villain,LSarma,DasSarma1992}. This result is confirmed in
Fig.~\ref{width} where the short time behavior is consistent with the growth
exponents $\beta=3/8$ in $d=1$ and $\beta=1/4$ in $d=2$ expected for the linear
MBE universality class~\cite{barabasi}. It is worth to mention that these models
may undergo crossovers to different universality classes in the asymptotic,
depending on the dimension and
model~\cite{wvbogo,Vvedensky,Xun,Chen2017,AaraoReis2004b,Punyindu}. The curves
in Fig.~\ref{width} are consistent with crossovers to different universality
classes at long times. One expects that DT is asymptotically consistent with the
non-linear MBE equation with $\lambda>0$~\cite{Xun,Luis2019,Luis2017}, for which
$\beta\approx 1/3$ and $1/5$ in $d=1$ and $d=2$, respectively\footnote{The
	exponents $\beta=1/3$ and $1/5$ are predictions of  the one-loop renormalization
	group~\cite{Villain,LSarma}. Two-loop calculations~\cite{Janssen}, however,
	predict corrections where the growth exponents are slightly smaller than these
	values.}, while crossovers to the Edwards-Wilkinson universality class with
$\beta=1/4$ in $d=1$ and $\beta=0$ (logarithmic growth) in $d=2$ are expected
for the WV model~\cite{wvbogo,Vvedensky}. The simulations with the kinetic barrier,
however, departs from the original dynamics after a transient which increases
with the diffusion of particles. For long times, an evolution consistent with
an uncorrelated growth 
{described by $\frac{\partial h}{\partial t}=\eta$, characterized
	by a growth exponent $\beta=1/2$~\cite{barabasi}, is observed.}
{This observation can be rationalized as follows. At long times, mounds
	interact weakly since the kinetic barrier reduces drastically inter-mound
	diffusion. Consider the idealized case of plateaus of size $L_0$ with
	an infinity barrier at their edges. A particle initially adsorbed on the top of
	a plateau will never slide down to its bottom. So, the probability that this
	plateau receives $R$ particles after one unity of time (deposition of $L$
	particles) is a binomial distribution
	\begin{equation}
	P(R)=\binom{L}{R} p^R (1-p)^{L-R}\simeq 
	\frac{1}{\sqrt{2\pi L_0}} e^{-\frac{(R-L_0)^2}{2L_0}},
	\label{eq:PR}
	\end{equation}
	where $p=L_0/L$ is the probability that a particle is deposited on this terrace
	and $1\ll L_0\ll L$ is assumed in the Gaussian limit in right-hand
	side of Eq.~\eqref{eq:PR}. We argue that this situation is similar to the weakly 
	interacting mound observed in our simulations.}

\begin{figure}[hbt]
	\centering
	\includegraphics[width=0.8\linewidth]{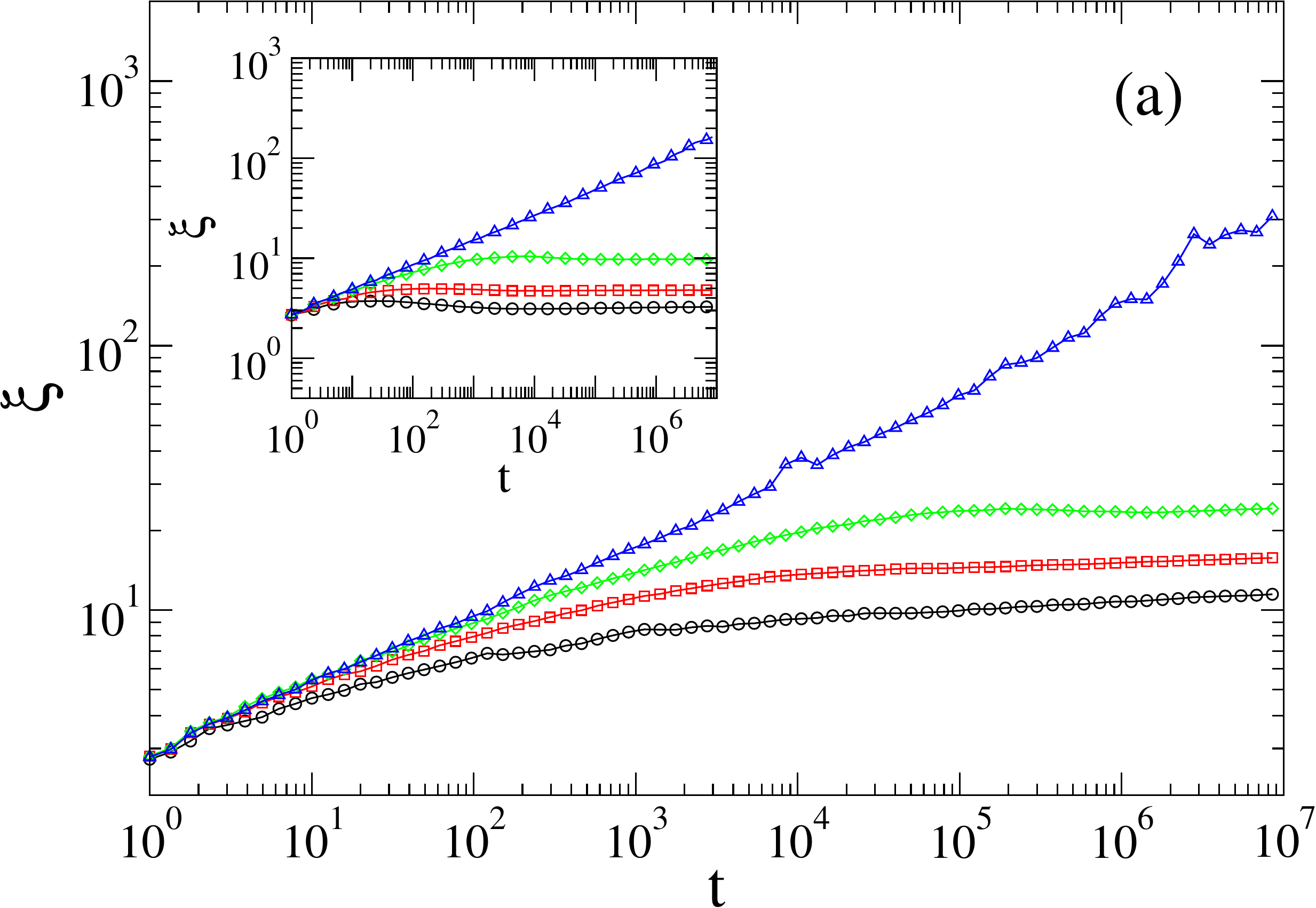}\\
	\includegraphics[width=0.8\linewidth]{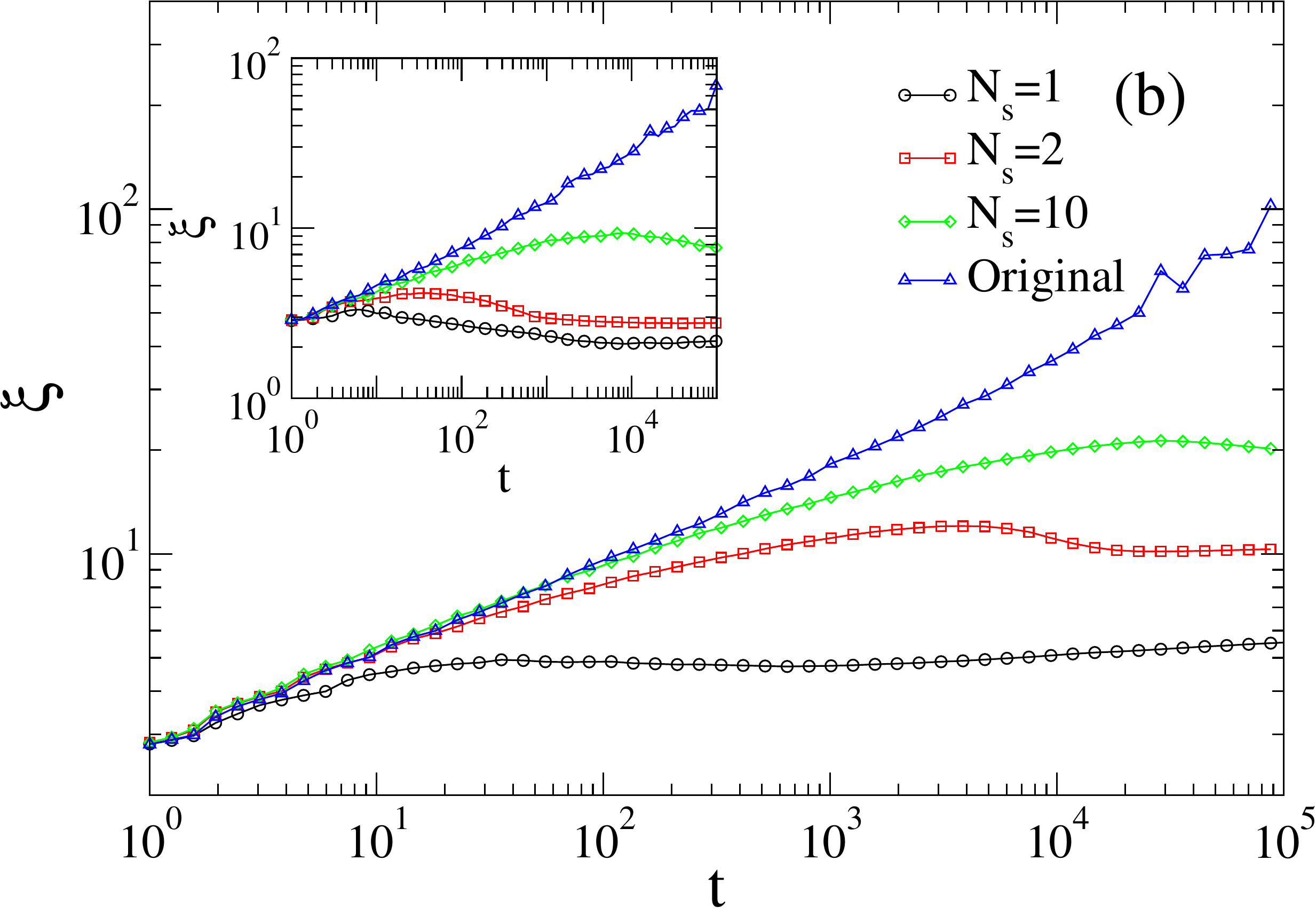}
	\caption{Characteristic length of mounds $\xi$ for WV (main plots) and DT
		(insets) models with and without the kinetic barrier in (a) one- and (b)
		two-dimensional substrates for different values of the parameter $N_s$ indicated
		in the legend.\label{xi}}
\end{figure}

In addition, as can be seen in Fig. \ref{xi}, the characteristic lateral lengths
of simulations with kinetic barrier saturate  after an initial transient in
values that increase with the parameter $N_s$ while the models without barrier
present coarsening with  $\xi\sim t^{1/z}$~\cite{barabasi}. The saturation
implies that the aspect ratio (height/width) of the mounds remains increasing
with time and the surface does not present slope selection forming columnar
growth. This property is also reflected in the asymptotic interface width
scaling as $w\sim t^{1/2}$. As explained previously, it can be interpreted as an
uncorrelated evolution of the columns, in which the 1/2 exponent comes out. The
results shown in the insets of Figs.~\ref{width} and \ref{xi} corroborate that
the DT model presents the same behavior of the WV model despite of the mounds
are less evident in the former.

\begin{figure}[ht]
	\centering
	\includegraphics[width=0.8\linewidth]{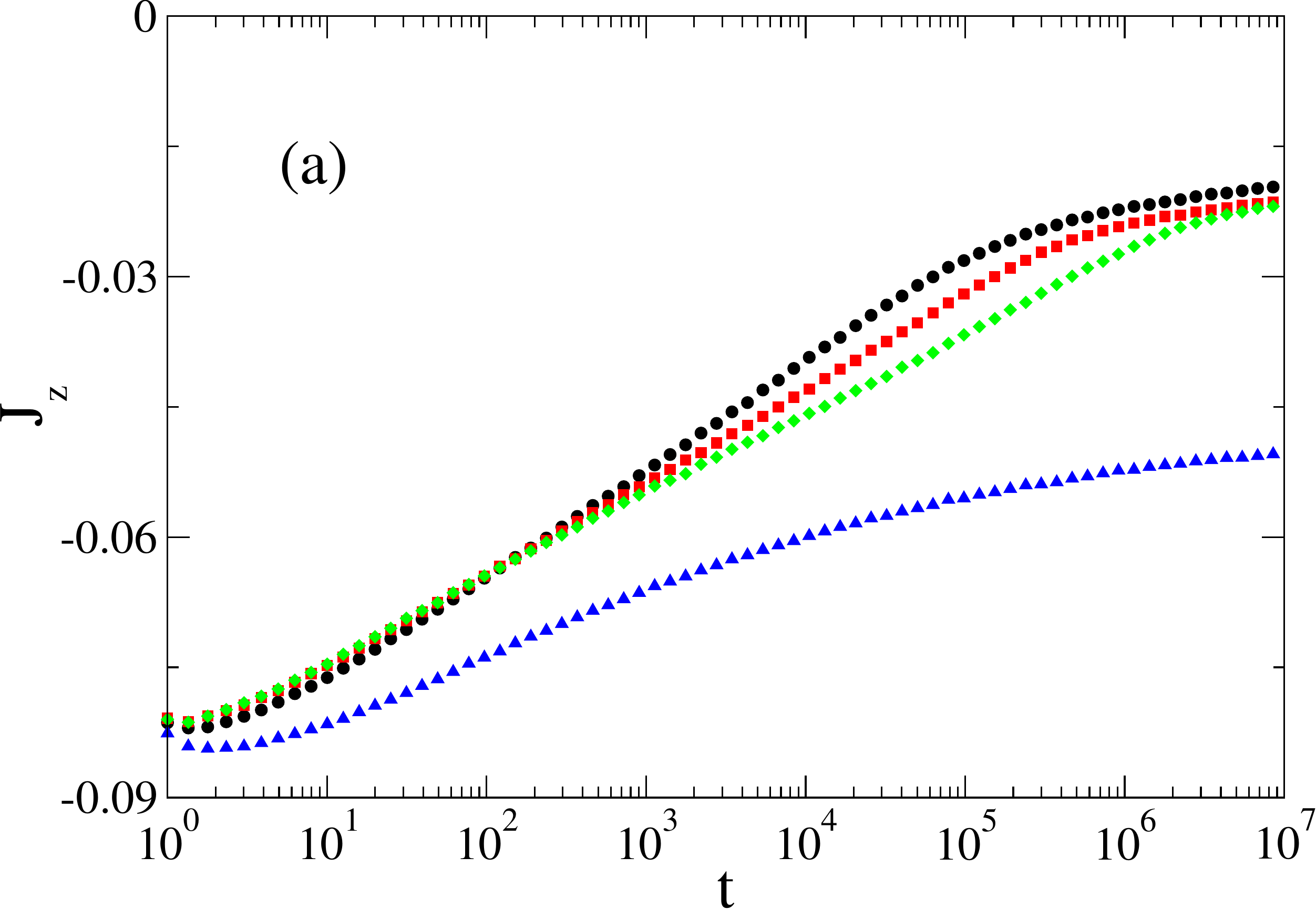}\\
	\includegraphics[width=0.8\linewidth]{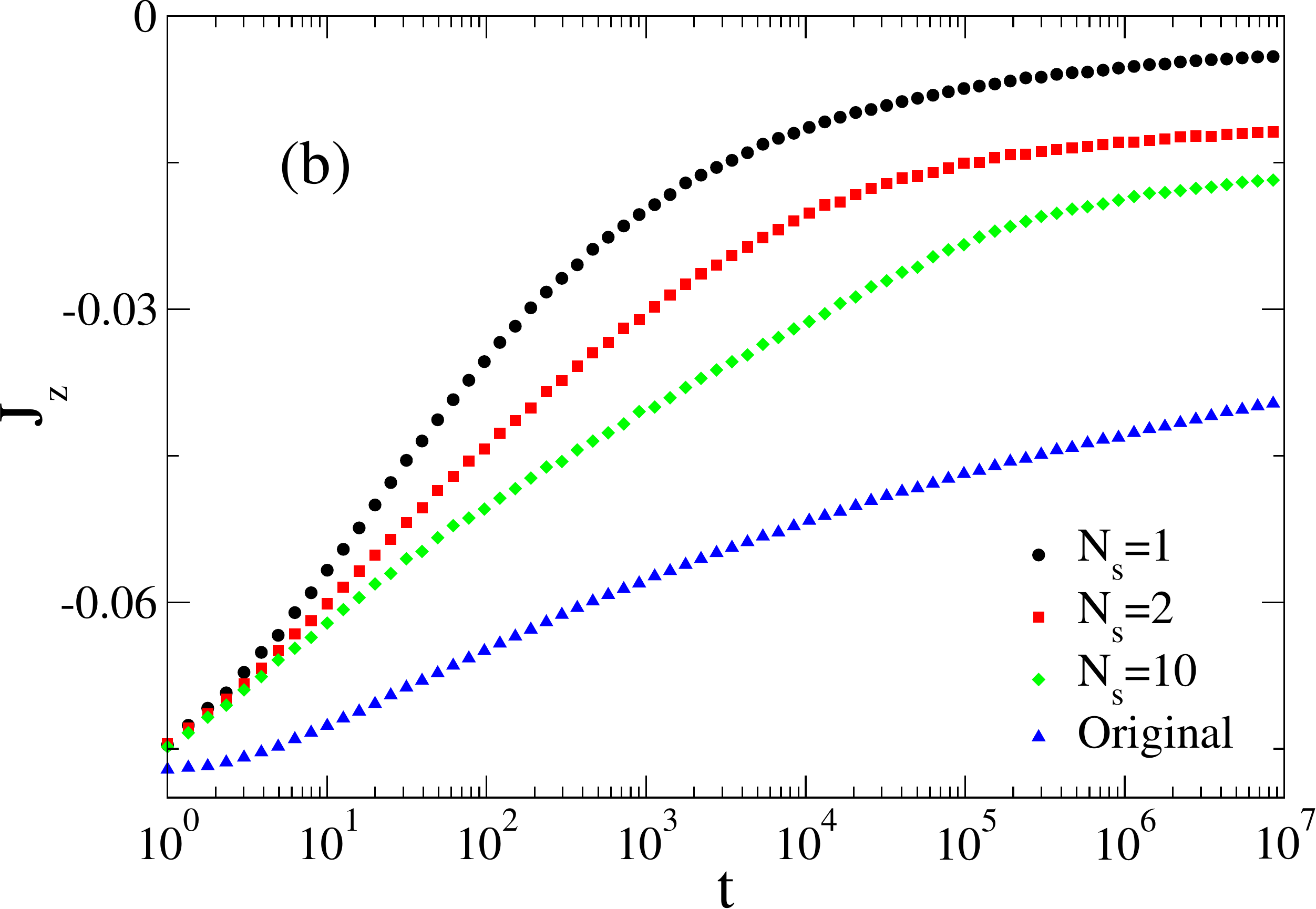}
	\caption{\label{current_1d} Evolution of the out-of-plane current for (a) WV and
		(b) DT models grown in one-dimensional substrates. Models with the kinetic
		barrier using $N_s =1$, $2$ and $10$ steps (indicated in the legend) 
		and the original version are shown.
		}
\end{figure}

Instability and mound formation can be investigated considering the surface
currents~\cite{Siegert1994,Krug1993}; see \cite{Krug1997} for details. In this
work, we  investigated the out-of-plane component of the current defined
as~\cite{Leal_Jstat}
\begin{equation}
J_z = \frac{1}{N}\sum_{(i,j)}  \mathrm{sgn}(\delta h) D(i,j) P_{\delta h}(i,j)
\end{equation}
where $\mathrm{sgn}(x)=1$ for $x>0$,  $\mathrm{sgn}(x)=-1$ for $x<0$, and $\mathrm{sgn}(0)=0$
is the definition of sign function,
$P_{\delta h}(i,j)$ is given by Eq.~\eqref{prob}, and $D(i,j)$ is the rate of
hopping attempts from site $i$ to $j$ and depend on the investigated model. The
sum runs over all $N$ pairs of nearest-neighbors of the lattice. Let $n_i$ be
the number of lateral bonds of site $i$ and $n^\text{max}_i$  the largest number
of bonds among the nearest-neighbors of $i$. For the WV model, $D(i,j)$ is given
by
\begin{equation}
D(i,j)=
\left\{
\begin{array}{cl}
1/q_i^\text{WV}, & \textrm{if } ~ n_j = n^\text{max}_i \textrm{  and  } n_i < n^\text{max}_i\\ 
0,  & \textrm{otherwise.}
\end{array}
\right.,
\end{equation} 
where $q_i^\text{WV}$ is the number of nearest-neighbors with $n^\text{max}_i$ lateral bonds.
We can express $D(i,j)$ for the DT as
\begin{equation}
D(i,j)=
\left\{
\begin{array}{cl}
1/q_i^\text{DT}, & \textrm{if } ~ n_j > 0  \textrm{ and } n_i = 0\\ 
0,  & \textrm{otherwise.}
\end{array}
\right.,
\end{equation}
where $q_i^{DT}$ is number of nearest-neighbors with at least one lateral bond.
The quantity $J_z$ is the average interlayer diffusion rate per site. 

\begin{table}[ht]
	\centering
	\extrarowsep=_20pt^3pt
	{\setlength{\tabcolsep}{8pt}
		\begin{tabular}{ccccc}
			
			\cline{2-5} 
			&\multicolumn{2}{ c}{$d=1$ } & \multicolumn{2}{ c}{ $d=2$ }  \\\cline{2-5}
			&WV& DT &    WV& DT \\\hline
			\multicolumn{1}{c}{  $N_s=1$ } & -0.0034     & -0.0042      & -0.015    &-5$\times 10^{-5}$  \\
			\multicolumn{1}{c}{  $N_s=2$ } & -0.0011     & -0.0052      & -0.019    & -3$\times 10^{-4}$  \\
			\multicolumn{1}{c}{  $N_s=10$} & -0.014      & -0.0053      & -0.020    & -5$\times 10^{-4}$ \\ 
			\multicolumn{1}{c}{  Original}    & -0.090       & -0.050       & -0.047     & -0.030 \\\hline
	\end{tabular}}
	\caption{\label{fit_parameters} Parameters $J_\infty$ obtained in
		the regression using the Eq.~(\ref{current_fit}) in the in the last decade
		of data of the out-plane current curves ($t>10^6$ for $d=1$ and $t>10^5$ for $d=2$).}
\end{table}

The currents for simulations in $d=1$ are presented in Fig.~\ref{current_1d}.
All versions in both $1+1$ and $2+1$ dimensions are characterized by a current
with a downward (negative) flux with the intensity decreasing  monotonically.
Considering the last decade of time, we estimated the current $J_\infty$ for
$t\rightarrow\infty$ using a regression with a simple allometric function in the form.
\begin{equation}
J_z = J_\infty + a t^{-\gamma}\label{current_fit},
\end{equation}
where $a$ and $\gamma$ are parameters. 
In all cases with step barrier, we obtained asymptotic small negative currents
with a non-universal value of $\gamma$. The results can be seen in
table~\ref{fit_parameters}. The currents for the standard models  are
considerably larger than in the cases  with barrier. The values for the DT model
with barrier are very small indicating that this current could be actually null
in the asymptotic limit as observed in thermally activated diffusion models
with ES step barriers~\cite{Leal_Jstat}. In the case of the WV model, the
current values may indicate the same asymptotic behavior, but our present
accuracy does not allow a conclusion on this issue.

\section{Conclusions}
\label{conclusion}

In this work, we investigate the effects of  a purely kinetic barrier caused by
the out-of-plane  step edge diffusion~\cite{Leal_JPCM}  on limited-mobility
growth models. The cases of studies were the benchmark models of
Wolf-Villain~\cite{WV} and Das Sarma-Tamborenea~\cite{DT}.  Large-scale
simulations were performed considering one- and two-dimensional substrates. It was
observed that the introduction of the kinetic barrier induces  the formation of
quasiregular mound structures differently from those obtained with the original
models that forms irregular (self-affine) structures in the interface. The
kinetic barrier stabilizes the mound width, leading to the formation of
quasiregular structures. The  interface width in models with kinetic barriers
has an initial regime similar to the  original models. However, a growth
exponent very close to $\beta = 1/2$ is observed for asymptotically long times.
Also, the characteristic lateral length saturates after a transient that
depends on the number of steps that an adatom can perform before irreversibly
stick in a position. These results are consistent  with mounds evolving
independently. The dynamics in both one- and two-dimensional substrates are
characterized by a strong reduction of downward current with respect to the  original
models. The  downward flux have an intensity decreasing  monotonically to a
asymptotic value that seems to be null for DT model and  small for WV, being the latter
possibly still subject to strong crossover effects in the present analysis.

A central contribution of this work is to show that a very simple mechanism
neglected in previous analysis, in which particles also diffuse in the direction
perpendicular to the substrate, is able to change markedly the surface morphology
of basic growth models with limited mobility.  Our results are qualitatively very similar to those
obtained when an explicit step barrier, with a smaller probability to move
downward, is considered~\cite{Rangdee}. Particularly, asymptotic mound
morphology has been reported for limited mobility models in $d=2$ without
barriers with the application of the noise reduction method~\cite{Chatraphorn2001}.
Our results corroborate this scenario since a small perturbation induces mound
instability in this kind of  processes while it alone does not produce mounds
in models with thermally activated diffusion~\cite{Leal_Jstat}. 

We expect that the concepts  investigated in this work will be applied to more
sophisticated models and aid the understanding of  pattern formation in film
growth and the  production of self-assembled structures for technological
applications.

\begin{acknowledgments}
SGA and SCF thank the financial support of Conselho Nacional de Desenvolvimento
Cient\'{i}fico e Tecnol\'{o}gico (CNPq) and Funda\c{c}\~{a}o de Pesquisa do
Estado de Minas Gerais (FAPEMIG). This study was financed in part by the
Coordena\c{c}\~{a}o de Aperfei\c{c}oamento de Pessoal de N\'{i}vel Superior -
Brasil (CAPES) - Finance Code 001.
\end{acknowledgments}


\begin{thebibliography}{51}%
	\makeatletter
	\providecommand \@ifxundefined [1]{%
		\@ifx{#1\undefined}
	}%
	\providecommand \@ifnum [1]{%
		\ifnum #1\expandafter \@firstoftwo
		\else \expandafter \@secondoftwo
		\fi
	}%
	\providecommand \@ifx [1]{%
		\ifx #1\expandafter \@firstoftwo
		\else \expandafter \@secondoftwo
		\fi
	}%
	\providecommand \natexlab [1]{#1}%
	\providecommand \enquote  [1]{``#1''}%
	\providecommand \bibnamefont  [1]{#1}%
	\providecommand \bibfnamefont [1]{#1}%
	\providecommand \citenamefont [1]{#1}%
	\providecommand \href@noop [0]{\@secondoftwo}%
	\providecommand \href [0]{\begingroup \@sanitize@url \@href}%
	\providecommand \@href[1]{\@@startlink{#1}\@@href}%
	\providecommand \@@href[1]{\endgroup#1\@@endlink}%
	\providecommand \@sanitize@url [0]{\catcode `\\12\catcode `\$12\catcode
		`\&12\catcode `\#12\catcode `\^12\catcode `\_12\catcode `\%12\relax}%
	\providecommand \@@startlink[1]{}%
	\providecommand \@@endlink[0]{}%
	\providecommand \url  [0]{\begingroup\@sanitize@url \@url }%
	\providecommand \@url [1]{\endgroup\@href {#1}{\urlprefix }}%
	\providecommand \urlprefix  [0]{URL }%
	\providecommand \Eprint [0]{\href }%
	\providecommand \doibase [0]{http://dx.doi.org/}%
	\providecommand \selectlanguage [0]{\@gobble}%
	\providecommand \bibinfo  [0]{\@secondoftwo}%
	\providecommand \bibfield  [0]{\@secondoftwo}%
	\providecommand \translation [1]{[#1]}%
	\providecommand \BibitemOpen [0]{}%
	\providecommand \bibitemStop [0]{}%
	\providecommand \bibitemNoStop [0]{.\EOS\space}%
	\providecommand \EOS [0]{\spacefactor3000\relax}%
	\providecommand \BibitemShut  [1]{\csname bibitem#1\endcsname}%
	\let\auto@bib@innerbib\@empty
	\bibitem [{\citenamefont {Michely}(2004)}]{michely2004islands}%
	\BibitemOpen
	\bibfield  {author} {\bibinfo {author} {\bibfnamefont {T.}~\bibnamefont
			{Michely}},\ }\href {https://www.springer.com/us/book/9783540407287} 
		{\emph {\bibinfo {title} {Islands, Mounds and
				Atoms}}}\ (\bibinfo  {publisher} {Springer Berlin Heidelberg Imprint
		Springer},\ \bibinfo {address} {Berlin, Heidelberg},\ \bibinfo {year}
	{2004})\BibitemShut {NoStop}%
	\bibitem [{\citenamefont {Evans}\ \emph {et~al.}(2006)\citenamefont {Evans},
		\citenamefont {Thiel},\ and\ \citenamefont {Bartelt}}]{Evans2006}%
	\BibitemOpen
	\bibfield  {author} {\bibinfo {author} {\bibfnamefont {J.}~\bibnamefont
			{Evans}}, \bibinfo {author} {\bibfnamefont {P.}~\bibnamefont {Thiel}}, \ and\
		\bibinfo {author} {\bibfnamefont {M.}~\bibnamefont {Bartelt}},\ }\bibfield
	{title} {\enquote {\bibinfo {title} {{Morphological evolution during
					epitaxial thin film growth: Formation of 2D islands and 3D mounds}},}\ }\href
	{http://linkinghub.elsevier.com/retrieve/pii/S0167572906000021} {\bibfield
		{journal} {\bibinfo  {journal} {Surf. Sci. Rep.}\ }\textbf {\bibinfo {volume}
			{61}},\ \bibinfo {pages} {1} (\bibinfo {year} {2006})}\BibitemShut {NoStop}%
	\bibitem [{\citenamefont {Barabasi}\ and\ \citenamefont
		{Stanley}(1995)}]{barabasi}%
	\BibitemOpen
	\bibfield  {author} {\bibinfo {author} {\bibfnamefont {A.-L.}\ \bibnamefont
			{Barabasi}}\ and\ \bibinfo {author} {\bibfnamefont {H.~E.}\ \bibnamefont
			{Stanley}},\ }\href{https://www.cambridge.org/core/books/fractal-concepts-in-surface-growth/0D9076FC287B60B2B1126BB165112F13} {\emph {\bibinfo {title} {Fractal Concepts in
				Surface Growth}}}\ (\bibinfo  {publisher} {Cambridge University Press},\
	\bibinfo {address} {Cambridge, England},\ \bibinfo {year} {1995})\BibitemShut
	{NoStop}%
	\bibitem [{\citenamefont {Meakin}(1998)}]{meakin}%
	\BibitemOpen
	\bibfield  {author} {\bibinfo {author} {\bibfnamefont {P.}~\bibnamefont
			{Meakin}},\ }\href{https://onlinelibrary.wiley.com/doi/abs/10.1002/%28SICI%291099-1204%28199911/12%2912%3A6%3C493%3A%3AAID-JNM346%3E3.0.CO%3B2-7} {\emph {\bibinfo {title} {Fractals, Scaling and
				Growth far from Equilibrium}}}\ (\bibinfo  {publisher} {Cambridge University
		Press},\ \bibinfo {address} {Cambridge, England},\ \bibinfo {year}
	{1998})\BibitemShut {NoStop}%
	\bibitem [{\citenamefont {Jorritsma}\ \emph {et~al.}(1997)\citenamefont
		{Jorritsma}, \citenamefont {Bijnagte}, \citenamefont {Rosenfeld},\ and\
		\citenamefont {Poelsema}}]{Jorritsma}%
	\BibitemOpen
	\bibfield  {author} {\bibinfo {author} {\bibfnamefont {L.~C.}\ \bibnamefont
			{Jorritsma}}, \bibinfo {author} {\bibfnamefont {M.}~\bibnamefont {Bijnagte}},
		\bibinfo {author} {\bibfnamefont {G.}~\bibnamefont {Rosenfeld}}, \ and\
		\bibinfo {author} {\bibfnamefont {B.}~\bibnamefont {Poelsema}},\ }\bibfield
	{title} {\enquote {\bibinfo {title} {Growth anisotropy and pattern formation
				in metal epitaxy},}\ }\href {\doibase 10.1103/PhysRevLett.78.911} {\bibfield
		{journal} {\bibinfo  {journal} {Phys. Rev. Lett.}\ }\textbf {\bibinfo
			{volume} {78}},\ \bibinfo {pages} {911} (\bibinfo {year} {1997})}\BibitemShut
	{NoStop}%
	\bibitem [{\citenamefont {Caspersen}\ \emph {et~al.}(2002)\citenamefont
		{Caspersen}, \citenamefont {Layson}, \citenamefont {Stoldt}, \citenamefont
		{Fournee}, \citenamefont {Thiel},\ and\ \citenamefont {Evans}}]{Caspersen}%
	\BibitemOpen
	\bibfield  {author} {\bibinfo {author} {\bibfnamefont {K.~J.}\ \bibnamefont
			{Caspersen}}, \bibinfo {author} {\bibfnamefont {A.~R.}\ \bibnamefont
			{Layson}}, \bibinfo {author} {\bibfnamefont {C.~R.}\ \bibnamefont {Stoldt}},
		\bibinfo {author} {\bibfnamefont {V.}~\bibnamefont {Fournee}}, \bibinfo
		{author} {\bibfnamefont {P.~A.}\ \bibnamefont {Thiel}}, \ and\ \bibinfo
		{author} {\bibfnamefont {J.~W.}\ \bibnamefont {Evans}},\ }\bibfield  {title}
	{\enquote {\bibinfo {title} {Development and ordering of mounds during
				metal(100) homoepitaxy},}\ }\href {\doibase 10.1103/PhysRevB.65.193407}
	{\bibfield  {journal} {\bibinfo  {journal} {Phys. Rev. B}\ }\textbf {\bibinfo
			{volume} {65}},\ \bibinfo {pages} {193407} (\bibinfo {year}
		{2002})}\BibitemShut {NoStop}%
	\bibitem [{\citenamefont {Han}\ \emph {et~al.}(2010)\citenamefont {Han},
		\citenamefont {\"Unal}, \citenamefont {Jing}, \citenamefont {Qin},
		\citenamefont {Jenks}, \citenamefont {Liu}, \citenamefont {Thiel},\ and\
		\citenamefont {Evans}}]{Han}%
	\BibitemOpen
	\bibfield  {author} {\bibinfo {author} {\bibfnamefont {Y.}~\bibnamefont
			{Han}}, \bibinfo {author} {\bibfnamefont {B.}~\bibnamefont {\"Unal}},
		\bibinfo {author} {\bibfnamefont {D.}~\bibnamefont {Jing}}, \bibinfo {author}
		{\bibfnamefont {F.}~\bibnamefont {Qin}}, \bibinfo {author} {\bibfnamefont
			{C.~J.}\ \bibnamefont {Jenks}}, \bibinfo {author} {\bibfnamefont {D.-J.}\
			\bibnamefont {Liu}}, \bibinfo {author} {\bibfnamefont {P.~A.}\ \bibnamefont
			{Thiel}}, \ and\ \bibinfo {author} {\bibfnamefont {J.~W.}\ \bibnamefont
			{Evans}},\ }\bibfield  {title} {\enquote {\bibinfo {title} {Formation and
				coarsening of Ag(110) bilayer islands on NiAl(110): Stm analysis and
				atomistic lattice-gas modeling},}\ }\href {\doibase
		10.1103/PhysRevB.81.115462} {\bibfield  {journal} {\bibinfo  {journal} {Phys.
				Rev. B}\ }\textbf {\bibinfo {volume} {81}},\ \bibinfo {pages} {115462}
		(\bibinfo {year} {2010})}\BibitemShut {NoStop}%
	\bibitem [{\citenamefont {Johnson}\ \emph {et~al.}(1994)\citenamefont
		{Johnson}, \citenamefont {Orme}, \citenamefont {Hunt}, \citenamefont {Graff},
		\citenamefont {Sudijono}, \citenamefont {Sander},\ and\ \citenamefont
		{Orr}}]{Johnson}%
	\BibitemOpen
	\bibfield  {author} {\bibinfo {author} {\bibfnamefont {M.~D.}\ \bibnamefont
			{Johnson}}, \bibinfo {author} {\bibfnamefont {C.}~\bibnamefont {Orme}},
		\bibinfo {author} {\bibfnamefont {A.~W.}\ \bibnamefont {Hunt}}, \bibinfo
		{author} {\bibfnamefont {D.}~\bibnamefont {Graff}}, \bibinfo {author}
		{\bibfnamefont {J.}~\bibnamefont {Sudijono}}, \bibinfo {author}
		{\bibfnamefont {L.~M.}\ \bibnamefont {Sander}}, \ and\ \bibinfo {author}
		{\bibfnamefont {B.~G.}\ \bibnamefont {Orr}},\ }\bibfield  {title} {\enquote
		{\bibinfo {title} {Stable and unstable growth in molecular beam epitaxy},}\
	}\href {\doibase 10.1103/PhysRevLett.72.116} {\bibfield  {journal} {\bibinfo
			{journal} {Phys. Rev. Lett.}\ }\textbf {\bibinfo {volume} {72}},\ \bibinfo
		{pages} {116} (\bibinfo {year} {1994})}\BibitemShut {NoStop}%
	\bibitem [{\citenamefont {Tadayyon-Eslami}\ \emph {et~al.}(2006)\citenamefont
		{Tadayyon-Eslami}, \citenamefont {Kan}, \citenamefont {Calhoun},\ and\
		\citenamefont {Phaneuf}}]{Tadayyon}%
	\BibitemOpen
	\bibfield  {author} {\bibinfo {author} {\bibfnamefont {T.}~\bibnamefont
			{Tadayyon-Eslami}}, \bibinfo {author} {\bibfnamefont {H.-C.}\ \bibnamefont
			{Kan}}, \bibinfo {author} {\bibfnamefont {L.~C.}\ \bibnamefont {Calhoun}}, \
		and\ \bibinfo {author} {\bibfnamefont {R.~J.}\ \bibnamefont {Phaneuf}},\
	}\bibfield  {title} {\enquote {\bibinfo {title} {Temperature-driven change in
				the unstable growth mode on patterned GaAs(001)},}\ }\href {\doibase
		10.1103/PhysRevLett.97.126101} {\bibfield  {journal} {\bibinfo  {journal}
			{Phys. Rev. Lett.}\ }\textbf {\bibinfo {volume} {97}},\ \bibinfo {pages}
		{126101} (\bibinfo {year} {2006})}\BibitemShut {NoStop}%
	\bibitem [{\citenamefont {Zorba}\ \emph {et~al.}(2006)\citenamefont {Zorba},
		\citenamefont {Shapir},\ and\ \citenamefont {Gao}}]{Zorba}%
	\BibitemOpen
	\bibfield  {author} {\bibinfo {author} {\bibfnamefont {S.}~\bibnamefont
			{Zorba}}, \bibinfo {author} {\bibfnamefont {Y.}~\bibnamefont {Shapir}}, \
		and\ \bibinfo {author} {\bibfnamefont {Y.}~\bibnamefont {Gao}},\ }\bibfield
	{title} {\enquote {\bibinfo {title} {Fractal-mound growth of pentacene thin
				films},}\ }\href {\doibase 10.1103/PhysRevB.74.245410} {\bibfield  {journal}
		{\bibinfo  {journal} {Phys. Rev. B}\ }\textbf {\bibinfo {volume} {74}},\
		\bibinfo {pages} {245410} (\bibinfo {year} {2006})}\BibitemShut {NoStop}%
	\bibitem [{\citenamefont {Hlawacek}\ \emph {et~al.}(2008)\citenamefont
		{Hlawacek}, \citenamefont {Puschnig}, \citenamefont {Frank}, \citenamefont
		{Winkler}, \citenamefont {Ambrosch-Draxl},\ and\ \citenamefont
		{Teichert}}]{Hlawacek}%
	\BibitemOpen
	\bibfield  {author} {\bibinfo {author} {\bibfnamefont {G.}~\bibnamefont
			{Hlawacek}}, \bibinfo {author} {\bibfnamefont {P.}~\bibnamefont {Puschnig}},
		\bibinfo {author} {\bibfnamefont {P.}~\bibnamefont {Frank}}, \bibinfo
		{author} {\bibfnamefont {A.}~\bibnamefont {Winkler}}, \bibinfo {author}
		{\bibfnamefont {C.}~\bibnamefont {Ambrosch-Draxl}}, \ and\ \bibinfo {author}
		{\bibfnamefont {C.}~\bibnamefont {Teichert}},\ }\bibfield  {title} {\enquote
		{\bibinfo {title} {Characterization of step-edge barriers in organic
				thin-film growth},}\ }\href {\doibase 10.1126/science.1159455} {\bibfield
		{journal} {\bibinfo  {journal} {Science}\ }\textbf {\bibinfo {volume}
			{321}},\ \bibinfo {pages} {108} (\bibinfo {year} {2008})}\BibitemShut
	{NoStop}%
	\bibitem [{\citenamefont {Ehrlich}\ and\ \citenamefont
		{Hudda}(1966)}]{Ehrlich}%
	\BibitemOpen
	\bibfield  {author} {\bibinfo {author} {\bibfnamefont {G.}~\bibnamefont
			{Ehrlich}}\ and\ \bibinfo {author} {\bibfnamefont {F.~G.}\ \bibnamefont
			{Hudda}},\ }\bibfield  {title} {\enquote {\bibinfo {title} {Atomic view of
				surface self-diffusion: Tungsten on tungsten},}\ }\href {\doibase
		10.1063/1.1726787} {\bibfield  {journal} {\bibinfo  {journal} {J.
				Chem. Phys.}\ }\textbf {\bibinfo {volume} {44}},\ \bibinfo {pages}
		{1039} (\bibinfo {year} {1966})}\BibitemShut {NoStop}%
	\bibitem [{\citenamefont {Schwoebel}\ and\ \citenamefont
		{Shipsey}(1966)}]{Schwoebel}%
	\BibitemOpen
	\bibfield  {author} {\bibinfo {author} {\bibfnamefont {R.~L.}\ \bibnamefont
			{Schwoebel}}\ and\ \bibinfo {author} {\bibfnamefont {E.~J.}\ \bibnamefont
			{Shipsey}},\ }\bibfield  {title} {\enquote {\bibinfo {title} {Step motion on
				crystal surfaces},}\ }\href {\doibase 10.1063/1.1707904} {\bibfield
		{journal} {\bibinfo  {journal} {J. Appl. Phys.}\ }\textbf
		{\bibinfo {volume} {37}},\ \bibinfo {pages} {3682} (\bibinfo {year}
		{1966})}\BibitemShut {NoStop}%
	\bibitem [{\citenamefont {Kanjanaput}\ \emph {et~al.}(2010)\citenamefont
		{Kanjanaput}, \citenamefont {Limkumnerd},\ and\ \citenamefont
		{Chatraphorn}}]{Kanjanaput2010}%
	\BibitemOpen
	\bibfield  {author} {\bibinfo {author} {\bibfnamefont {W.}~\bibnamefont
			{Kanjanaput}}, \bibinfo {author} {\bibfnamefont {S.}~\bibnamefont
			{Limkumnerd}}, \ and\ \bibinfo {author} {\bibfnamefont {P.}~\bibnamefont
			{Chatraphorn}},\ }\bibfield  {title} {\enquote {\bibinfo {title} {{Growth
					instability due to lattice-induced topological currents in limited-mobility
					epitaxial growth models}},}\ }\href
	{http://link.aps.org/doi/10.1103/PhysRevE.82.041607} {\bibfield  {journal}
		{\bibinfo  {journal} {Phys. Rev. E}\ }\textbf {\bibinfo {volume} {82}},\
		\bibinfo {pages} {041607} (\bibinfo {year} {2010})}\BibitemShut {NoStop}%
	\bibitem [{\citenamefont {Murty}\ and\ \citenamefont
		{Cooper}(2003)}]{Murty2003}%
	\BibitemOpen
	\bibfield  {author} {\bibinfo {author} {\bibfnamefont {M.~R.}\ \bibnamefont
			{Murty}}\ and\ \bibinfo {author} {\bibfnamefont {B.}~\bibnamefont {Cooper}},\
	}\bibfield  {title} {\enquote {\bibinfo {title} {Influence of step edge
				diffusion on surface morphology during epitaxy},}\ }\href@noop {} {\bibfield
		{journal} {\bibinfo  {journal} {Surf. Sci.}\ }\textbf {\bibinfo {volume}
			{539}},\ \bibinfo {pages} {91} (\bibinfo {year} {2003})}\BibitemShut
	{NoStop}%
	\bibitem [{\citenamefont {Pierre-Louis}\ \emph {et~al.}(1999)\citenamefont
		{Pierre-Louis}, \citenamefont {D'Orsogna},\ and\ \citenamefont
		{Einstein}}]{Pierre-Louis1999}%
	\BibitemOpen
	\bibfield  {author} {\bibinfo {author} {\bibfnamefont {O.}~\bibnamefont
			{Pierre-Louis}}, \bibinfo {author} {\bibfnamefont {M.~R.}\ \bibnamefont
			{D'Orsogna}}, \ and\ \bibinfo {author} {\bibfnamefont {T.~L.}\ \bibnamefont
			{Einstein}},\ }\bibfield  {title} {\enquote {\bibinfo {title} {{Edge
					Diffusion during Growth: The Kink Ehrlich-Schwoebel Effect and Resulting
					Instabilities}},}\ }\href
	{https://link.aps.org/doi/10.1103/PhysRevLett.82.3661} {\bibfield  {journal}
		{\bibinfo  {journal} {Phys. Rev. Lett.}\ }\textbf {\bibinfo {volume} {82}},\
		\bibinfo {pages} {3661} (\bibinfo {year} {1999})}\BibitemShut {NoStop}%
	\bibitem [{\citenamefont {Yang}\ \emph {et~al.}(2007)\citenamefont {Yang},
		\citenamefont {Sun}, \citenamefont {Zhang},\ and\ \citenamefont
		{Jia}}]{Yang}%
	\BibitemOpen
	\bibfield  {author} {\bibinfo {author} {\bibfnamefont {H.}~\bibnamefont
			{Yang}}, \bibinfo {author} {\bibfnamefont {Q.}~\bibnamefont {Sun}}, \bibinfo
		{author} {\bibfnamefont {Z.}~\bibnamefont {Zhang}}, \ and\ \bibinfo {author}
		{\bibfnamefont {Y.}~\bibnamefont {Jia}},\ }\bibfield  {title} {\enquote
		{\bibinfo {title} {Upward self-diffusion of adatoms and small clusters on
				facets of fcc metal (110) surfaces},}\ }\href
	{https://link.aps.org/doi/10.1103/PhysRevB.76.115417} {\bibfield  {journal}
		{\bibinfo  {journal} {Phys. Rev. B}\ }\textbf {\bibinfo {volume} {76}},\
		\bibinfo {pages} {115417} (\bibinfo {year} {2007})}\BibitemShut {NoStop}%
	\bibitem [{\citenamefont {Clarke}\ and\ \citenamefont
		{Vvedensky}(1987)}]{CV_PRL}%
	\BibitemOpen
	\bibfield  {author} {\bibinfo {author} {\bibfnamefont {S.}~\bibnamefont
			{Clarke}}\ and\ \bibinfo {author} {\bibfnamefont {D.~D.}\ \bibnamefont
			{Vvedensky}},\ }\bibfield  {title} {\enquote {\bibinfo {title} {Origin of
				reflection high-energy electron-diffraction intensity oscillations during
				molecular-beam epitaxy: A computational modeling approach},}\ }\href
	{\doibase 10.1103/PhysRevLett.58.2235} {\bibfield  {journal} {\bibinfo
			{journal} {Phys. Rev. Lett.}\ }\textbf {\bibinfo {volume} {58}},\ \bibinfo
		{pages} {2235} (\bibinfo {year} {1987})}\BibitemShut {NoStop}%
	\bibitem [{\citenamefont {Clarke}\ and\ \citenamefont
		{Vvedensky}(1988)}]{Clarke1988}%
	\BibitemOpen
	\bibfield  {author} {\bibinfo {author} {\bibfnamefont {S.}~\bibnamefont
			{Clarke}}\ and\ \bibinfo {author} {\bibfnamefont {D.~D.}\ \bibnamefont
			{Vvedensky}},\ }\bibfield  {title} {\enquote {\bibinfo {title} {{Growth
					kinetics and step density in reflection high-energy electron diffraction
					during molecular-beam epitaxy}},}\ }\href
	{http://scitation.aip.org/content/aip/journal/jap/63/7/10.1063/1.341041}
	{\bibfield  {journal} {\bibinfo  {journal} {J. Appl. Phys.}\ }\textbf
		{\bibinfo {volume} {63}},\ \bibinfo {pages} {2272} (\bibinfo {year}
		{1988})}\BibitemShut {NoStop}%
	\bibitem [{\citenamefont {Leal}\ \emph
		{et~al.}(2011{\natexlab{a}})\citenamefont {Leal}, \citenamefont {Ferreira},\
		and\ \citenamefont {Ferreira}}]{Leal_JPCM}%
	\BibitemOpen
	\bibfield  {author} {\bibinfo {author} {\bibfnamefont {F.~F.}\ \bibnamefont
			{Leal}}, \bibinfo {author} {\bibfnamefont {S.~C.}\ \bibnamefont {Ferreira}},
		\ and\ \bibinfo {author} {\bibfnamefont {S.~O.}\ \bibnamefont {Ferreira}},\
	}\bibfield  {title} {\enquote {\bibinfo {title} {{Modelling of epitaxial film
					growth with an Ehrlich-Schwoebel barrier dependent on the step height}},}\
	}\href
	{http://stacks.iop.org/0953-8984/23/i=29/a=292201?key=crossref.9247eb5fac27043f6f6da8363d2658c5}
	{\bibfield  {journal} {\bibinfo  {journal} {J. Phys. Condens. Matter}\
		}\textbf {\bibinfo {volume} {23}},\ \bibinfo {pages} {292201} (\bibinfo
		{year} {2011}{\natexlab{a}})}\BibitemShut {NoStop}%
	\bibitem [{\citenamefont {Wolf}\ and\ \citenamefont {Villain}(1990)}]{WV}%
	\BibitemOpen
	\bibfield  {author} {\bibinfo {author} {\bibfnamefont {D.~E.}\ \bibnamefont
			{Wolf}}\ and\ \bibinfo {author} {\bibfnamefont {J.}~\bibnamefont {Villain}},\
	}\bibfield  {title} {\enquote {\bibinfo {title} {Growth with surface
				diffusion},}\ }\href{https://iopscience.iop.org/article/10.1209/0295-5075/13/5/002} {\bibfield  {journal} {\bibinfo  {journal} {Eur.
				Lett.}\ }\textbf {\bibinfo {volume} {13}},\ \bibinfo {pages} {389} (\bibinfo
		{year} {1990})}\BibitemShut {NoStop}%
	\bibitem [{\citenamefont {Das~Sarma}\ and\ \citenamefont
		{Tamborenea}(1991)}]{DT}%
	\BibitemOpen
	\bibfield  {author} {\bibinfo {author} {\bibfnamefont {S.}~\bibnamefont
			{Das~Sarma}}\ and\ \bibinfo {author} {\bibfnamefont {P.}~\bibnamefont
			{Tamborenea}},\ }\bibfield  {title} {\enquote {\bibinfo {title} {A new
				universality class for kinetic growth: One-dimensional molecular-beam
				epitaxy},}\ }\href {\doibase 10.1103/PhysRevLett.66.325} {\bibfield
		{journal} {\bibinfo  {journal} {Phys. Rev. Lett.}\ }\textbf {\bibinfo
			{volume} {66}},\ \bibinfo {pages} {325} (\bibinfo {year} {1991})}\BibitemShut
	{NoStop}%
	\bibitem [{\citenamefont {\ifmmode~\check{S}\else \v{S}\fi{}milauer}\ and\
		\citenamefont {Kotrla}(1994)}]{Smilauer}%
	\BibitemOpen
	\bibfield  {author} {\bibinfo {author} {\bibfnamefont {P.}~\bibnamefont
			{\ifmmode~\check{S}\else \v{S}\fi{}milauer}}\ and\ \bibinfo {author}
		{\bibfnamefont {M.}~\bibnamefont {Kotrla}},\ }\bibfield  {title} {\enquote
		{\bibinfo {title} {Crossover effects in the Wolf-Villain model of epitaxial
				growth in 1+1 and 2+1 dimensions},}\ }\href {\doibase
		10.1103/PhysRevB.49.5769} {\bibfield  {journal} {\bibinfo  {journal} {Phys.
				Rev. B}\ }\textbf {\bibinfo {volume} {49}},\ \bibinfo {pages} {5769}
		(\bibinfo {year} {1994})}\BibitemShut {NoStop}%
	\bibitem [{\citenamefont {P\ifmmode~\check{r}\else \v{r}\fi{}edota}\ and\
		\citenamefont {Kotrla}(1996)}]{Milan}%
	\BibitemOpen
	\bibfield  {author} {\bibinfo {author} {\bibfnamefont {M.}~\bibnamefont
			{P\ifmmode~\check{r}\else \v{r}\fi{}edota}}\ and\ \bibinfo {author}
		{\bibfnamefont {M.}~\bibnamefont {Kotrla}},\ }\bibfield  {title} {\enquote
		{\bibinfo {title} {Stochastic equations for simple discrete models of
				epitaxial growth},}\ }\href {\doibase 10.1103/PhysRevE.54.3933} {\bibfield
		{journal} {\bibinfo  {journal} {Phys. Rev. E}\ }\textbf {\bibinfo {volume}
			{54}},\ \bibinfo {pages} {3933} (\bibinfo {year} {1996})}\BibitemShut
	{NoStop}%
	\bibitem [{\citenamefont {Huang}\ and\ \citenamefont {Gu}(1996)}]{Huang}%
	\BibitemOpen
	\bibfield  {author} {\bibinfo {author} {\bibfnamefont {Z.-F.}\ \bibnamefont
			{Huang}}\ and\ \bibinfo {author} {\bibfnamefont {B.-L.}\ \bibnamefont {Gu}},\
	}\bibfield  {title} {\enquote {\bibinfo {title} {Growth equations for the
				Wolf-Villain and Das Sarma-Tamborenea models of molecular-beam epitaxy},}\
	}\href {\doibase 10.1103/PhysRevE.54.5935} {\bibfield  {journal} {\bibinfo
			{journal} {Phys. Rev. E}\ }\textbf {\bibinfo {volume} {54}},\ \bibinfo
		{pages} {5935} (\bibinfo {year} {1996})}\BibitemShut {NoStop}%
	\bibitem [{\citenamefont {Haselwandter}\ and\ \citenamefont
		{Vvedensky}(2007)}]{HaselwandterPRL}%
	\BibitemOpen
	\bibfield  {author} {\bibinfo {author} {\bibfnamefont {C.~A.}\ \bibnamefont
			{Haselwandter}}\ and\ \bibinfo {author} {\bibfnamefont {D.~D.}\ \bibnamefont
			{Vvedensky}},\ }\bibfield  {title} {\enquote {\bibinfo {title} {Multiscale
				theory of fluctuating interfaces: Renormalization of atomistic models},}\
	}\href {\doibase 10.1103/PhysRevLett.98.046102} {\bibfield  {journal}
		{\bibinfo  {journal} {Phys. Rev. Lett.}\ }\textbf {\bibinfo {volume} {98}},\
		\bibinfo {pages} {046102} (\bibinfo {year} {2007})}\BibitemShut {NoStop}%
	\bibitem [{\citenamefont {Haselwandter}\ and\ \citenamefont
		{Vvedensky}(2008)}]{HaselwandterPRE}%
	\BibitemOpen
	\bibfield  {author} {\bibinfo {author} {\bibfnamefont {C.~A.}\ \bibnamefont
			{Haselwandter}}\ and\ \bibinfo {author} {\bibfnamefont {D.~D.}\ \bibnamefont
			{Vvedensky}},\ }\bibfield  {title} {\enquote {\bibinfo {title}
			{Renormalization of stochastic lattice models: Epitaxial surfaces},}\ }\href
	{\doibase 10.1103/PhysRevE.77.061129} {\bibfield  {journal} {\bibinfo
			{journal} {Phys. Rev. E}\ }\textbf {\bibinfo {volume} {77}},\ \bibinfo
		{pages} {061129} (\bibinfo {year} {2008})}\BibitemShut {NoStop}%
	\bibitem [{\citenamefont {Das~Sarma}\ \emph {et~al.}(2002)\citenamefont
		{Das~Sarma}, \citenamefont {Chatraphorn},\ and\ \citenamefont
		{Toroczkai}}]{Sarma}%
	\BibitemOpen
	\bibfield  {author} {\bibinfo {author} {\bibfnamefont {S.}~\bibnamefont
			{Das~Sarma}}, \bibinfo {author} {\bibfnamefont {P.~P.}\ \bibnamefont
			{Chatraphorn}}, \ and\ \bibinfo {author} {\bibfnamefont {Z.}~\bibnamefont
			{Toroczkai}},\ }\bibfield  {title} {\enquote {\bibinfo {title} {Universality
				class of discrete solid-on-solid limited mobility nonequilibrium growth
				models for kinetic surface roughening},}\ }\href {\doibase
		10.1103/PhysRevE.65.036144} {\bibfield  {journal} {\bibinfo  {journal} {Phys.
				Rev. E}\ }\textbf {\bibinfo {volume} {65}},\ \bibinfo {pages} {036144}
		(\bibinfo {year} {2002})}\BibitemShut {NoStop}%
	\bibitem [{\citenamefont {Punyindu}\ and\ \citenamefont
		{Das~Sarma}(1998)}]{Punyindu}%
	\BibitemOpen
	\bibfield  {author} {\bibinfo {author} {\bibfnamefont {P.}~\bibnamefont
			{Punyindu}}\ and\ \bibinfo {author} {\bibfnamefont {S.}~\bibnamefont
			{Das~Sarma}},\ }\bibfield  {title} {\enquote {\bibinfo {title} {Noise
				reduction and universality in limited-mobility models of nonequilibrium
				growth},}\ }\href {\doibase 10.1103/PhysRevE.57.R4863} {\bibfield  {journal}
		{\bibinfo  {journal} {Phys. Rev. E}\ }\textbf {\bibinfo {volume} {57}},\
		\bibinfo {pages} {R4863} (\bibinfo {year} {1998})}\BibitemShut {NoStop}%
	\bibitem [{\citenamefont {Alves}\ and\ \citenamefont {Moreira}(2011)}]{wvbogo}%
	\BibitemOpen
	\bibfield  {author} {\bibinfo {author} {\bibfnamefont {S.~G.}\ \bibnamefont
			{Alves}}\ and\ \bibinfo {author} {\bibfnamefont {J.~G.}\ \bibnamefont
			{Moreira}},\ }\bibfield  {title} {\enquote {\bibinfo {title} {Transitions in
				a probabilistic interface growth model},}\ }\href{https://iopscience.iop.org/article/10.1088/1742-5468/2011/04/P04022/meta} {\bibfield
		{journal} {\bibinfo  {journal} {J. Stat. Mech.: Theory Exp.}\ }\textbf {\bibinfo {volume} {2011}},\ \bibinfo {pages} {P04022}
		(\bibinfo {year} {2011})}\BibitemShut {NoStop}%
	\bibitem [{\citenamefont {Xun}\ \emph {et~al.}(2012)\citenamefont {Xun},
		\citenamefont {Tang}, \citenamefont {Han}, \citenamefont {Xia}, \citenamefont
		{Hao},\ and\ \citenamefont {Li}}]{Xun}%
	\BibitemOpen
	\bibfield  {author} {\bibinfo {author} {\bibfnamefont {Z.}~\bibnamefont
			{Xun}}, \bibinfo {author} {\bibfnamefont {G.}~\bibnamefont {Tang}}, \bibinfo
		{author} {\bibfnamefont {K.}~\bibnamefont {Han}}, \bibinfo {author}
		{\bibfnamefont {H.}~\bibnamefont {Xia}}, \bibinfo {author} {\bibfnamefont
			{D.}~\bibnamefont {Hao}}, \ and\ \bibinfo {author} {\bibfnamefont
			{Y.}~\bibnamefont {Li}},\ }\bibfield  {title} {\enquote {\bibinfo {title}
			{Asymptotic dynamic scaling behavior of the (1+1)-dimensional Wolf-Villain
				model},}\ }\href {\doibase 10.1103/PhysRevE.85.041126} {\bibfield  {journal}
		{\bibinfo  {journal} {Phys. Rev. E}\ }\textbf {\bibinfo {volume} {85}},\
		\bibinfo {pages} {041126} (\bibinfo {year} {2012})}\BibitemShut {NoStop}%
	\bibitem [{\citenamefont {Luis}\ \emph {et~al.}(2019)\citenamefont {Luis},
		\citenamefont {de~Assis}, \citenamefont {Ferreira},\ and\ \citenamefont
		{Andrade}}]{Luis2019}%
	\BibitemOpen
	\bibfield  {author} {\bibinfo {author} {\bibfnamefont {{Edwin~E.~Mozo}}\
			\bibnamefont {Luis}}, \bibinfo {author} {\bibfnamefont {T.~A.}\ \bibnamefont
			{de~Assis}}, \bibinfo {author} {\bibfnamefont {S.~C.}\ \bibnamefont
			{Ferreira}}, \ and\ \bibinfo {author} {\bibfnamefont {R.~F.~S.}\ \bibnamefont
			{Andrade}},\ }\bibfield  {title} {\enquote {\bibinfo {title} {{Local
					roughness exponent in the nonlinear molecular-beam-epitaxy universality class
					in one dimension}},}\ }\href
	{https://link.aps.org/doi/10.1103/PhysRevE.99.022801} {\bibfield  {journal}
		{\bibinfo  {journal} {Phys. Rev. E}\ }\textbf {\bibinfo {volume} {99}},\
		\bibinfo {pages} {022801} (\bibinfo {year} {2019})}\BibitemShut {NoStop}%
	\bibitem [{\citenamefont {Aar\~ao Reis}(2010)}]{Aarao2010}%
	\BibitemOpen
	\bibfield  {author} {\bibinfo {author} {\bibfnamefont {F.~D.~A.}\
			\bibnamefont {Aar\~ao Reis}},\ }\bibfield  {title} {\enquote {\bibinfo
			{title} {Dynamic scaling in thin-film growth with irreversible step-edge
				attachment},}\ }\href {\doibase 10.1103/PhysRevE.81.041605} {\bibfield
		{journal} {\bibinfo  {journal} {Phys. Rev. E}\ }\textbf {\bibinfo {volume}
			{81}},\ \bibinfo {pages} {041605} (\bibinfo {year} {2010})}\BibitemShut
	{NoStop}%
	\bibitem [{\citenamefont {Aar\~ao Reis}(2013)}]{Aarao2013}%
	\BibitemOpen
	\bibfield  {author} {\bibinfo {author} {\bibfnamefont {F.~D.~A.}\
			\bibnamefont {Aar\~ao Reis}},\ }\bibfield  {title} {\enquote {\bibinfo
			{title} {Normal dynamic scaling in the class of the nonlinear
				molecular-beam-epitaxy equation},}\ }\href {\doibase
		10.1103/PhysRevE.88.022128} {\bibfield  {journal} {\bibinfo  {journal} {Phys.
				Rev. E}\ }\textbf {\bibinfo {volume} {88}},\ \bibinfo {pages} {022128}
		(\bibinfo {year} {2013})}\BibitemShut {NoStop}%
	\bibitem [{\citenamefont {To}\ \emph {et~al.}(2018)\citenamefont {To},
		\citenamefont {de~Sousa},\ and\ \citenamefont {{Aar{\~{a}}o Reis}}}]{To2018}%
	\BibitemOpen
	\bibfield  {author} {\bibinfo {author} {\bibfnamefont {T.~B.}\ \bibnamefont
			{To}}, \bibinfo {author} {\bibfnamefont {V.~B.}\ \bibnamefont {de~Sousa}}, \
		and\ \bibinfo {author} {\bibfnamefont {F.~D.}\ \bibnamefont {{Aar{\~{a}}o
					Reis}}},\ }\bibfield  {title} {\enquote {\bibinfo {title} {{Thin film growth
					models with long surface diffusion lengths}},}\ }\href
	{https://linkinghub.elsevier.com/retrieve/pii/S0378437118308884} {\bibfield
		{journal} {\bibinfo  {journal} {Phys. A Stat. Mech. its Appl.}\ }\textbf
		{\bibinfo {volume} {511}},\ \bibinfo {pages} {240} (\bibinfo {year}
		{2018})}\BibitemShut {NoStop}%
	\bibitem [{\citenamefont {Rangdee}\ and\ \citenamefont
		{Chatraphorn}(2006)}]{Rangdee}%
	\BibitemOpen
	\bibfield  {author} {\bibinfo {author} {\bibfnamefont {R.}~\bibnamefont
			{Rangdee}}\ and\ \bibinfo {author} {\bibfnamefont {P.}~\bibnamefont
			{Chatraphorn}},\ }\bibfield  {title} {\enquote {\bibinfo {title} {Effects of
				the Ehrlich-Schwoebel potential barrier on the Wolf-Villain model simulations
				for thin film growth},}\ }\href {\doibase 10.1016/j.susc.2005.12.021}
	{\bibfield  {journal} {\bibinfo  {journal} {Surf. Science}\ }\textbf
		{\bibinfo {volume} {600}},\ \bibinfo {pages} {914 } (\bibinfo {year}
		{2006})}\BibitemShut {NoStop}%
	\bibitem [{\citenamefont {Sarma}\ and\ \citenamefont
		{Punyindu}(1999)}]{DasSarma_SC}%
	\BibitemOpen
	\bibfield  {author} {\bibinfo {author} {\bibfnamefont {S.~D.}\ \bibnamefont
			{Sarma}}\ and\ \bibinfo {author} {\bibfnamefont {P.}~\bibnamefont
			{Punyindu}},\ }\bibfield  {title} {\enquote {\bibinfo {title} {A discrete
				model for non-equilibrium growth under surface diffusion bias},}\ }\href
	{\doibase 10.1016/S0039-6028(99)00209-5} {\bibfield  {journal} {\bibinfo
			{journal} {Surf. Science}\ }\textbf {\bibinfo {volume} {424}},\ \bibinfo
		{pages} {L339 } (\bibinfo {year} {1999})}\BibitemShut {NoStop}%
	\bibitem [{\citenamefont {Chatraphorn}\ \emph {et~al.}(2001)\citenamefont
		{Chatraphorn}, \citenamefont {Toroczkai},\ and\ \citenamefont {{Das
				Sarma}}}]{Chatraphorn2001}%
	\BibitemOpen
	\bibfield  {author} {\bibinfo {author} {\bibfnamefont {{P. Punyindu}}~ \bibnamefont{Chatraphorn}}, 
		                      \bibinfo {author} {\bibfnamefont {Z.}~\bibnamefont{Toroczkai}}, \ and\
	                          \bibinfo {author} {\bibfnamefont {S.}~\bibnamefont {{Das Sarma}}},\ }
                          \bibfield  {title} {\enquote {\bibinfo {title} {{Epitaxial
					mounding in limited-mobility models of surface growth}},}\ }\href
	{http://link.aps.org/doi/10.1103/PhysRevB.64.205407} {\bibfield  {journal}
		{\bibinfo  {journal} {Phys. Rev. B}\ }\textbf {\bibinfo {volume} {64}},\
		\bibinfo {pages} {205407} (\bibinfo {year} {2001})}\BibitemShut {NoStop}%
	\bibitem [{\citenamefont {Vvedensky}(2003)}]{Vvedensky}%
	\BibitemOpen
	\bibfield  {author} {\bibinfo {author} {\bibfnamefont {D.~D.}\ \bibnamefont
			{Vvedensky}},\ }\bibfield  {title} {\enquote {\bibinfo {title} {Crossover and
				universality in the Wolf-Villain model},}\ }\href {\doibase
		10.1103/PhysRevE.68.010601} {\bibfield  {journal} {\bibinfo  {journal} {Phys.
				Rev. E}\ }\textbf {\bibinfo {volume} {68}},\ \bibinfo {pages} {010601}
		(\bibinfo {year} {2003})}\BibitemShut {NoStop}%
	\bibitem [{\citenamefont {{J. Villain}}(1991)}]{Villain}%
	\BibitemOpen
	\bibfield  {author} {\bibinfo {author} {\bibnamefont {{J. Villain}}},\
	}\bibfield  {title} {\enquote {\bibinfo {title} {Continuum models of crystal
				growth from atomic beams with and without desorption},}\ }\href {\doibase
		10.1051/jp1:1991114} {\bibfield  {journal} {\bibinfo  {journal} {J. Phys. I
				France}\ }\textbf {\bibinfo {volume} {1}},\ \bibinfo {pages} {19} (\bibinfo
		{year} {1991})}\BibitemShut {NoStop}%
	\bibitem [{\citenamefont {Lai}\ and\ \citenamefont {Das~Sarma}(1991)}]{LSarma}%
	\BibitemOpen
	\bibfield  {author} {\bibinfo {author} {\bibfnamefont {Z.-W.}\ \bibnamefont
			{Lai}}\ and\ \bibinfo {author} {\bibfnamefont {S.}~\bibnamefont
			{Das~Sarma}},\ }\bibfield  {title} {\enquote {\bibinfo {title} {Kinetic
				growth with surface relaxation: Continuum versus atomistic models},}\ }\href
	{http://link.aps.org/doi/10.1103/PhysRevLett.66.2348} {\bibfield  {journal}
		{\bibinfo  {journal} {Phys. Rev. Lett.}\ }\textbf {\bibinfo {volume} {66}},\
		\bibinfo {pages} {2348} (\bibinfo {year} {1991})}\BibitemShut {NoStop}%
	\bibitem [{\citenamefont {El-Shehawey}(2000)}]{Shehawey}%
	\BibitemOpen
	\bibfield  {author} {\bibinfo {author} {\bibfnamefont {M.~A.}\ \bibnamefont
			{El-Shehawey}},\ }\bibfield  {title} {\enquote {\bibinfo {title} {Absorption
				probabilities for a random walk between two partially absorbing boundaries:
				I},}\ }\href{https://iopscience.iop.org/article/10.1088/0305-4470/33/49/301} {\bibfield  {journal} {\bibinfo  {journal} {J.
				Phys. A: Math. Gen.}\ }\textbf {\bibinfo {volume} {33}},\
		\bibinfo {pages} {9005} (\bibinfo {year} {2000})}\BibitemShut {NoStop}%
	\bibitem [{\citenamefont {{Das Sarma}}\ and\ \citenamefont
		{Ghaisas}(1992)}]{DasSarma1992}%
	\BibitemOpen
	\bibfield  {author} {\bibinfo {author} {\bibfnamefont {S.}~\bibnamefont {{Das
					Sarma}}}\ and\ \bibinfo {author} {\bibfnamefont {S.~V.}\ \bibnamefont
			{Ghaisas}},\ }\bibfield  {title} {\enquote {\bibinfo {title} {{Solid-on-solid
					rules and models for nonequilibrium growth in 2+1 dimensions}},}\ }\href
	{\doibase 10.1103/PhysRevLett.69.3762} {\bibfield  {journal} {\bibinfo
			{journal} {Phys. Rev. Lett.}\ }\textbf {\bibinfo {volume} {69}},\ \bibinfo
		{pages} {3762} (\bibinfo {year} {1992})}\BibitemShut {NoStop}%
	\bibitem [{\citenamefont {Chen}\ \emph {et~al.}(2017)\citenamefont {Chen},
		\citenamefont {Tang}, \citenamefont {Xun}, \citenamefont {Zhu},\ and\
		\citenamefont {Zhang}}]{Chen2017}%
	\BibitemOpen
	\bibfield  {author} {\bibinfo {author} {\bibfnamefont {Y.}~\bibnamefont
			{Chen}}, \bibinfo {author} {\bibfnamefont {G.}~\bibnamefont {Tang}}, \bibinfo
		{author} {\bibfnamefont {Z.}~\bibnamefont {Xun}}, \bibinfo {author}
		{\bibfnamefont {L.}~\bibnamefont {Zhu}}, \ and\ \bibinfo {author}
		{\bibfnamefont {Z.}~\bibnamefont {Zhang}},\ }\bibfield  {title} {\enquote
		{\bibinfo {title} {{Schramm-Loewner evolution theory of the asymptotic
					behaviors of (2+1)-dimensional Wolf-Villain model}},}\ }\href
	{http://dx.doi.org/10.1016/j.physa.2016.08.057
		http://linkinghub.elsevier.com/retrieve/pii/S037843711630574X} {\bibfield
		{journal} {\bibinfo  {journal} {Phys. A Stat. Mech. its Appl.}\ }\textbf
		{\bibinfo {volume} {465}},\ \bibinfo {pages} {613} (\bibinfo {year}
		{2017})}\BibitemShut {NoStop}%
	\bibitem [{\citenamefont {{Aar{\~{a}}o Reis}}(2004)}]{AaraoReis2004b}%
	\BibitemOpen
	\bibfield  {author} {\bibinfo {author} {\bibfnamefont {{F.~D.~A.~Aar{\~{a}}o}}\
			\bibnamefont {{Reis}}},\ }\bibfield  {title} {\enquote {\bibinfo
			{title} {{Numerical study of discrete models in the class of the nonlinear
					Molecular Beam Epitaxy equation}},}\ }\href
	{https://link.aps.org/doi/10.1103/PhysRevE.70.031607} {\bibfield  {journal}
		{\bibinfo  {journal} {Phys. Rev. E}\ }\textbf {\bibinfo {volume} {70}},\
		\bibinfo {pages} {031607} (\bibinfo {year} {2004})}\BibitemShut {NoStop}%
	\bibitem [{\citenamefont {Luis}\ \emph {et~al.}(2017)\citenamefont {Luis},
		\citenamefont {de~Assis},\ and\ \citenamefont {Ferreira}}]{Luis2017}%
	\BibitemOpen
	\bibfield  {author} {\bibinfo {author} {\bibfnamefont {{Edwin~E.~Mozo}}\
			\bibnamefont {Luis}}, \bibinfo {author} {\bibfnamefont {T.~A.}\ \bibnamefont
			{de~Assis}}, \ and\ \bibinfo {author} {\bibfnamefont {S.~C.}\ \bibnamefont
			{Ferreira}},\ }\bibfield  {title} {\enquote {\bibinfo {title} {{Optimal
					detrended fluctuation analysis as a tool for the determination of the
					roughness exponent of the mounded surfaces}},}\ }\href
	{http://link.aps.org/doi/10.1103/PhysRevE.95.042801} {\bibfield  {journal}
		{\bibinfo  {journal} {Phys. Rev. E}\ }\textbf {\bibinfo {volume} {95}},\
		\bibinfo {pages} {042801} (\bibinfo {year} {2017})}\BibitemShut {NoStop}%
	\bibitem [{\citenamefont {Janssen}(1997)}]{Janssen}%
	\BibitemOpen
	\bibfield  {author} {\bibinfo {author} {\bibfnamefont {H.~K.}\ \bibnamefont
			{Janssen}},\ }\bibfield  {title} {\enquote {\bibinfo {title} {On critical
				exponents and the renormalization of the coupling constant in growth models
				with surface diffusion},}\ }\href
	{http://link.aps.org/doi/10.1103/PhysRevLett.78.1082} {\bibfield  {journal}
		{\bibinfo  {journal} {Phys. Rev. Lett.}\ }\textbf {\bibinfo {volume} {78}},\
		\bibinfo {pages} {1082} (\bibinfo {year} {1997})}\BibitemShut {NoStop}%
	\bibitem [{\citenamefont {Siegert}\ and\ \citenamefont
		{Plischke}(1994)}]{Siegert1994}%
	\BibitemOpen
	\bibfield  {author} {\bibinfo {author} {\bibfnamefont {M.}~\bibnamefont
			{Siegert}}\ and\ \bibinfo {author} {\bibfnamefont {M.}~\bibnamefont
			{Plischke}},\ }\bibfield  {title} {\enquote {\bibinfo {title} {{Slope
					Selection and Coarsening in Molecular Beam Epitaxy}},}\ }\href
	{http://link.aps.org/doi/10.1103/PhysRevLett.73.1517} {\bibfield  {journal}
		{\bibinfo  {journal} {Phys. Rev. Lett.}\ }\textbf {\bibinfo {volume} {73}},\
		\bibinfo {pages} {1517} (\bibinfo {year} {1994})}\BibitemShut {NoStop}%
	\bibitem [{\citenamefont {Krug}\ \emph {et~al.}(1993)\citenamefont {Krug},
		\citenamefont {Plischke},\ and\ \citenamefont {Siegert}}]{Krug1993}%
	\BibitemOpen
	\bibfield  {author} {\bibinfo {author} {\bibfnamefont {J.}~\bibnamefont
			{Krug}}, \bibinfo {author} {\bibfnamefont {M.}~\bibnamefont {Plischke}}, \
		and\ \bibinfo {author} {\bibfnamefont {M.}~\bibnamefont {Siegert}},\
	}\bibfield  {title} {\enquote {\bibinfo {title} {{Surface diffusion currents
					and the universality classes of growth}},}\ }\href
	{https://link.aps.org/doi/10.1103/PhysRevLett.70.3271} {\bibfield  {journal}
		{\bibinfo  {journal} {Phys. Rev. Lett.}\ }\textbf {\bibinfo {volume} {70}},\
		\bibinfo {pages} {3271} (\bibinfo {year} {1993})}\BibitemShut {NoStop}%
	\bibitem [{\citenamefont {Krug}(1997)}]{Krug1997}%
	\BibitemOpen
	\bibfield  {author} {\bibinfo {author} {\bibfnamefont {J.}~\bibnamefont
			{Krug}},\ }\bibfield  {title} {\enquote {\bibinfo {title} {{Origins of scale
					invariance in growth processes}},}\ }\href
	{http://www.tandfonline.com/doi/abs/10.1080/00018739700101498} {\bibfield
		{journal} {\bibinfo  {journal} {Adv. Phys.}\ }\textbf {\bibinfo {volume}
			{46}},\ \bibinfo {pages} {139} (\bibinfo {year} {1997})}\BibitemShut
	{NoStop}%
	\bibitem [{\citenamefont {Leal}\ \emph
		{et~al.}(2011{\natexlab{b}})\citenamefont {Leal}, \citenamefont {Oliveira},\
		and\ \citenamefont {Ferreira}}]{Leal_Jstat}%
	\BibitemOpen
	\bibfield  {author} {\bibinfo {author} {\bibfnamefont {F.~F.}\ \bibnamefont
			{Leal}}, \bibinfo {author} {\bibfnamefont {T.~J.}\ \bibnamefont {Oliveira}},
		\ and\ \bibinfo {author} {\bibfnamefont {S.~C.}\ \bibnamefont {Ferreira}},\
	}\bibfield  {title} {\enquote {\bibinfo {title} {{Kinetic modelling of
					epitaxial film growth with up- and downward step barriers}},}\ }\href
	{http://stacks.iop.org/1742-5468/2011/i=09/a=P09018?key=crossref.72e1a82e5e15653a11bd7c44b2902efa}
	{\bibfield  {journal} {\bibinfo  {journal} {J. Stat. Mech. Theory Exp.}\
		}\textbf {\bibinfo {volume} {2011}},\ \bibinfo {pages} {P09018} (\bibinfo
		{year} {2011}{\natexlab{b}})}\BibitemShut {NoStop}%
\end{thebibliography}

%


\end{document}